\documentclass[twocolumn,showpacs,preprintnumbers,amsmath,amssymb,superscriptaddress]{revtex4-2} 
\usepackage{chemformula} 
\usepackage{graphicx}
\usepackage[colorlinks=true,citecolor=blue,urlcolor=blue,linkcolor=blue,bookmarksopen]{hyperref}
\usepackage{color}
\usepackage{braket}
\usepackage{physics}
\usepackage{gensymb}
\usepackage{soul}
\usepackage[T1]{fontenc}

\begin{document}

\title{Topological Jackiw-Rebbi States in Photonic Van der Waals Heterostructures}

\author{Sam A. Randerson}
\email{s.a.randerson@sheffield.ac.uk}
\affiliation{School of Mathematical and Physical Sciences, The University of Sheffield, Sheffield S3 7RH, U.K.} 
\author{Paul Bouteyre}
\thanks{S.A. Randerson and P. Bouteyre contributed equally to this work as first authors.}
\affiliation{School of Mathematical and Physical Sciences, The University of Sheffield, Sheffield S3 7RH, U.K.}
\author{Xuerong Hu}
\affiliation{School of Mathematical and Physical Sciences, The University of Sheffield, Sheffield S3 7RH, U.K.}
\author{Oscar J. Palma Chaundler}
\affiliation{School of Mathematical and Physical Sciences, The University of Sheffield, Sheffield S3 7RH, U.K.} 
\author{Alexander J. Knight}
\affiliation{School of Mathematical and Physical Sciences, The University of Sheffield, Sheffield S3 7RH, U.K.}
\author{Helgi Sigur{\dh}sson}
\affiliation{Science Institute, University of Iceland, Dunhagi 3, IS-107 Reykjavik, Iceland}
\affiliation{Institute of Experimental Physics, Faculty of Physics, University of Warsaw, ul. Pasteura 5, PL-02-093 Warsaw, Poland}
\author{Casey K. Cheung}
\affiliation{Department of Physics and Astronomy, The University of Manchester, Oxford Road, Manchester, M13 9PL, U.K.}
\affiliation{National Graphene Institute, The University of Manchester, Oxford Road, Manchester, M13 9PL, U.K.}
\author{Yue Wang}
\affiliation{School of Physics Engineering and Technology, University of York, York YO10 5DD, U.K.}
\author{Kenji Watanabe}
\affiliation{Research Center for Electronic and Optical Materials, National Institute for Materials Science, 1-1 Namiki, Tsukuba, 305-0044 Japan}
\author{Takashi Taniguchi}
\affiliation{Research Center for Materials Nanoarchitectonics, National Institute for Materials Science, 1-1 Namiki, Tsukuba, 305-0044 Japan}
\author{Roman Gorbachev}
\affiliation{Department of Physics and Astronomy, The University of Manchester, Oxford Road, Manchester, M13 9PL, U.K.}
\affiliation{National Graphene Institute, The University of Manchester, Oxford Road, Manchester, M13 9PL, U.K.}
\author{Alexander I. Tartakovskii}
\email{a.tartakovskii@sheffield.ac.uk}
\affiliation{School of Mathematical and Physical Sciences, The University of Sheffield, Sheffield S3 7RH, U.K.} 

\date{\today}

\begin{abstract}

Topological phenomena, first studied in solid state physics, have seen increased interest for applications in nanophotonics owing to highly controllable light confinement with inherent robustness to defects. Photonic crystals can be designed to host topologically protected interface states for directional light transport, localization and robust lasing via tuning of the bulk topological invariant. At the same time, van der Waals (vdW) materials, in both their monolayer and quasi-bulk forms, are emerging as exciting additions to the field of nanophotonics, with a range of unique optoelectronic properties and intrinsic adherence to any type of host material, allowing fabrication of complex multi-layer structures. We present here a 1D topological photonic platform made from stacked nanostructured and planar layers of quasi-bulk \ch{WS2} to achieve Jackiw-Rebbi (JR) interface states between two topologically distinct gratings in the near-infrared range around $750$ nm. Such states are measured in the far-field with angle-resolved reflectance contrast measurements, exhibiting linewidth of $10$ meV and highly directional emission with an angular bandwidth of $8.0\degree$. Subsequent local mapping of the structure via sub-wavelength resolution scattering-type scanning near-field optical microscopy (s-SNOM) reveals strong spatial confinement of the JR state to the grating interface region. Finally, we couple in the JR state the photoluminescence of monolayer \ch{WSe_2} incorporated in a five-layer vdW grating heterostructure, giving rise to directional enhancement of the excitonic emission of up to $22$ times that of uncoupled monolayer, thus demonstrating the potential of the topological interface states for highly directional light emission in addition to light scattering.

\end{abstract}


\maketitle

\section{Introduction}
\label{sec:intro}

Topological photonics is a rapidly evolving field promising intrinsically robust control of light at the nanoscale compared to conventional nanophotonic systems, where performance varies greatly with fabrication quality. Stemming from the discovery of topological insulators in 1980 via the quantum Hall effect ~\cite{klitzing1980new}, optical analogues have been more recently investigated in the form of topologically-protected photonic interface states, for efficient waveguiding and light manipulation with inherent immunity to back-scatter ~\cite{wang2009observation}. For instance, 2D spin Hall photonic lattices have been realised using silicon, yielding confined interface states with unidirectional propagation selected by the polarisation handedness of the incident light ~\cite{Shalaev2019,parappurath2020}. Such topologically-protected resonances were measured in the far-field, but can also be directly probed in the near-field via scanning near-field optical microscopy (SNOM) methods ~\cite{Arora2021,orsini2024deep}. 

Of closer importance to this work, one dimensional topological photonic crystals ~\cite{Ozawa_RMP2019} have been studied to realise highly localised interface states between systems of different Zak phases ~\cite{zak1989berry}. Such states manifest within the leaky radiation continuum as Jackiw-Rebbi (JR) interface states ~\cite{Jackiw1976}, characterised by robust and directional emission situated at the centre of the photonic band gap in energy. The history of JR states can be traced back to a solution to the 1D Dirac equation, where fermions of opposing mass form highly localised interface states regardless of small changes in the masses ~\cite{Jackiw1976}. Via comparison to the Su-Schrieffer-Heeger (SSH) model ~\cite{heeger1988solitons}, photonic JR interface states have been identified between grating structures with opposing Zak phases in simulation ~\cite{Lee2021,Lee2022} and experiment ~\cite{An2024,orsini2024deep,Wang_arxiv2025,choi2025topological}, as well as in dielectric particle arrays ~\cite{st2017lasing,Gorlach_PRB2019}. Such JR interface states boast applications such as lasing ~\cite{st2017lasing,Ota2018} and beam shaping ~\cite{Lee2022,choi2025topological}, therefore providing additional, topologically-protected channels for advanced control of light on the nanoscale.

Layered 2D crystals, known as van der Waals (vdW) materials, have also garnered significant interest over the last two decades owing to their inherent interlayer attractive forces, and unique optoelectronic properties that undergo large changes between bulk and monolayer ~\cite{wang2012electronics,mak2016photonics}. For example, monolayers of transition metal dichalcogenides (TMDs), a prominent member of the vdW family, exhibit direct band gaps with high oscillator strength excitons, yielding efficient emission ~\cite{splendiani2010emerging,mak2010atomically}. Their bulk counterparts are emerging as capable and versatile candidates for sub-wavelength light confinement and control ~\cite{zotev2023van,randerson2024high}, owing to high refractive indices with uniaxial anisotropy, and low losses over a large portion of the visible wavelength range ~\cite{munkhbat2022optical}. TMDs and other vdW materials can be readily fabricated by hand via mechanical exfoliation, adhering easily to a wide range of substrates without the need for chemical bonding or lattice matching. This allows for incorporation of TMD monolayers onto silicon-based structures for example, achieving spin Hall topological interface states ~\cite{Li2021} and optical valley Hall currents ~\cite{Lundt2019} for exciton-polaritons. Furthermore, standard lithography and etching techniques can be successfully applied to vdW materials to realise high quality patterned structures in experiment ~\cite{Munkhbat2020,munkhbat2023nanostructured,danielsen2021super} such as 2D photonic crystals etched from bulk TMDs, observed to host unidirectional propagating interface states ~\cite{isoniemi2024realization}. Van der Waals materials thus present exciting additions to nanophotonics and light-matter physics, aiding where specific optoelectronic properties are required, or advanced multi-layer heterostructures need to be built.

We note that current experimental studies on JR states in 1D gratings primarily consider low index materials such as \ch{SiN} for the grating device, hence requiring large structures of several microns in thickness and near micron periods to operate within the near-IR range ~\cite{choi2025topological}. More compact structures fabricated from low index \ch{TiO2} suffer from greatly reduced contrast of the mode dispersions, and require additional ion-assisted deposition steps ~\cite{Wang_arxiv2025}. Here, we achieve high contrast grating modes and topological JR interface states in much more compact structures ($<100$ nm thickness), fabricated using high index \ch{WS2} ($n_\mathrm{WS_2}=4$ ~\cite{munkhbat2022optical}) via simple mechanical exfoliation and $``$pick-and-place$"$ polymer stamp transfer techniques of up to five individual planar layers, to operate within the near-IR range around $750$ nm.

\begin{figure*}
\centering
\includegraphics[width=\linewidth]{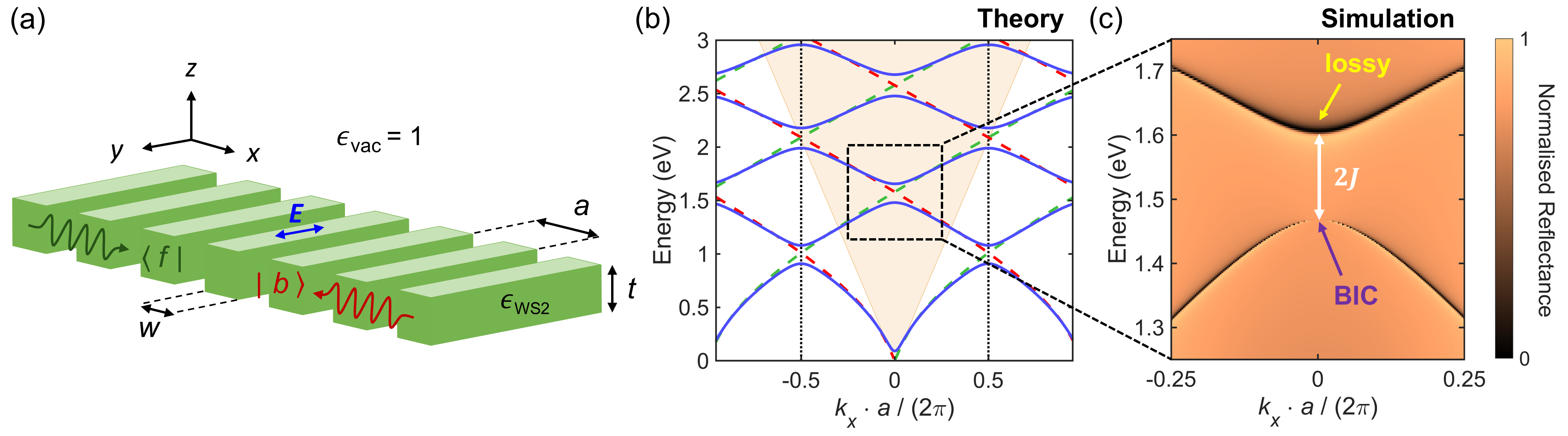}
  \caption{\textbf{Theoretical model of coupled grating modes.} \textbf{(a)} Schematic of suspended \ch{WS2} grating with parameters $a$, $t$, and $w$ corresponding to the grating period, thickness, and width of each \ch{WS2} wire respectively. Incident plane wave is polarised parallel to the grooves (i.e. along $y$) denoted by the blue double arrow. $\bra{f}$ and $\ket{b}$ correspond to the forwards and backwards propagating grating modes respectively, travelling perpendicular to the grooves. \textbf{(b)} Scheme of the dispersion replicas of forward and backwards propagating guided modes denoted by the green and red dashed lines respectively. The first Brillouin zone is denoted by the dotted black lines. Modes are band folded into the first Brillouin zone and exist within the light cone (orange shaded region). Blue curves correspond to the energies of diffractively coupled guided modes, showing gap opening at branch crossings. \textbf{(c)} Normalised angle-resolved reflectance simulation of a suspended \ch{WS2} grating showing emergence of a lossy mode and infinitely narrow linewidth BIC on the upper and lower energy branches respectively.}
  \label{Figure1}
\end{figure*}

In this work we begin by outlining the analytical model of our system, defining the Hamiltonian of propagating modes within a 1D grating to achieve a bound state in the continuum (BIC) and lossy mode, split by a band gap at the $\Gamma$-point in momentum space (see Figure \ref{Figure1}). Topological effects of the band structures are further considered via comparison to the 1D Dirac equation, where combining gratings with opposite topological phases results in the formation of a localized JR interface state. We then detail the design and rationale behind our inverted gratings, where an additional high index \ch{WS_2} slab is simulated on top of a \ch{WS_2} grating to promote band inversion and a topological phase transition upon tuning the filling factor (FF). Leveraging on the inherent interlayer attractive forces of TMDs, we realise such inverted gratings via polymer-stamp transfer of bulk \ch{WS_2} flakes onto pre-fabricated double grating devices. Through this, we experimentally demonstrate precise tuning of the photonic band gap and band inversion via the control of the refractive index contrast in the grating akin to Refs. ~\cite{lee2019band,Lee2021}, with subsequent observation of photonic JR states between topologically distinct structures in the visible-to-near-infrared regime. The far-field responses of all resonances are fully characterised with angle-resolved reflectance contrast throughout momentum space, with the JR state exhibiting an angular bandwidth of $8.0\degree$, along with a $10$ meV linewidth in energy. We verify the strong spatial localisation of the JR state by direct probing via scattering-type SNOM (s-SNOM). Finally, photoluminescence from an hBN-encapsulated \ch{WSe_2} monolayer within a five-layer double grating heterostructure is coupled to the JR interface state, yielding highly directional emission up to $22$ times stronger than uncoupled monolayer.



\section{Theoretical model}
\label{sec:model}

\subsection{Guided modes of the photonic grating}
Assuming continuous translational symmetry in the lateral $y$ direction, the periodic modulation of the grating acts on the guided photonic modes through a potential operator of the form $V = u(x) w(z)$, where $u(x) = u(x+a)$ is along the grating period and $w(z)$ is a step function along the vertical direction as illustrated schematically in Figure \ref{Figure1} (a). Counter-propagating guided electromagnetic modes along $x$ that differ by an integer number of the primitive reciprocal lattice number $K = 2 \pi/a$ are diffractively coupled and exhibit Bragg reflection. Replicas of the guided mode dispersion fold across the Brillouin zone with a gap opening around their crossing point as in Figure \ref{Figure1} (b). For example, second order diffractive coupling between modes $|e^{\pm iqx}\rangle$ with the wavevectors $q = \pm K + k_x$, where $k_x \ll K$, opens a gap at the $\Gamma$-point which corresponds to a second-order stop band. Here, we will focus on the transverse electric (TE) modes, whose dispersion is approximately given by $\omega_{\pm K} = \omega_K \pm v k$ with the group velocity $v$ around $q=K$. As the guided modes $|e^{\pm iKx}\rangle$ fold into the Brillouin zone, they also couple with lossy modes residing at normal incidence $q=0$ within the light cone (shaded orange region) through a first order diffraction process ~\cite{Sigurdsson_Nanopho2024}. Sufficiently close to the crossing point (see boxed region of Figure \ref{Figure1} (b)) the dynamics of photons in the grating can be described by a non-Hermitian Dirac-like Hamiltonian ~\cite{Lu_PhotRes2020, Lee2021, Sigurdsson_Nanopho2024},
\begin{equation} \label{eq.phoH}
    \hat{H} = \begin{pmatrix}
        v k_x & Je^{i \phi} \\ 
        Je^{-i \phi} & -vk_x
    \end{pmatrix}
    - i \gamma \begin{pmatrix}
        1 & 1 \\
        1 & 1
    \end{pmatrix}.
\end{equation}
Here, we have set $\omega_K = 0$ without loss of generality. The first term contains the Hermitian diffractive coupling parameter $J e^{i \phi}$ between counter-propagating modes resulting in a gap of size $2J$, as in Figure \ref{Figure1} (c). The second anti-Hermitian term describes coupling $\gamma$ of the modes to the lossy radiative continuum through a Friedrich-Wintgen type process ~\cite{Friedrich_PRA1985} (see Supplementary Note 1 for more details).

The mirror symmetry of the grating, taken to be at $x=0$ so that $u(x) = u(-x)$, implies that $\phi \in \{0,\pi\}$, and guarantees the presence of a symmetry protected photonic BIC at the $\Gamma$-point in the antisymmetric energy branch ~\cite{Azzam_AdvOptMat2021}. The BIC is also topologically tied to a polarization vortex in the far-field ~\cite{Doeleman2018}. The phase $\phi$ of the diffractive coupling mechanism can be flipped from $0 \to \pi$ by adjusting the pitch and filling factor of the grating ~\cite{Lu_PhotRes2020, Lee2021, Sigurdsson_Nanopho2024}. Physically, this phase flip corresponds to inverting the energy hierarchy between the symmetric and antisymmetric standing wave Bloch states. 

The eigenvalues of Eqn.~\eqref{eq.phoH} describe the dispersion of the guided photons at low momenta, corresponding to the angle-resolved reflectance simulation (see Methods) shown in Figure \ref{Figure1} (c) using the Rigorous Coupled-Wave Analysis (RCWA) technique,
\begin{equation} \label{eq.pho}
\omega_{\pm}(k_x) =  - i \gamma \pm \sqrt{(vk_x)^2 + J^2 - \gamma^2 -2 iJ \gamma  \cos{(\phi)}}.
\end{equation}
The eigenvectors are derived in Supplementary Note 1. Here, $\omega_\pm$ refer to the upper ($+$) and lower ($-$) energy branches (bands) of the grating standing waves. The diffractive coupling $J\gg\gamma$ is the main parameter responsible for the bandgap opening at $k_x=0$. When $\phi = 0$, a BIC mode of an infinite lifetime appears in the center of the lower $\omega_-$ antisymmetric branch while a lossy $2 \gamma$ mode appears in the upper symmetric $\omega_+$ branch (see Figure \ref{Figure1} (c)). If $\phi = \pi$, equivalent to $J \to -J$, band inversion takes place and the antisymmetric state containing the BIC is now higher in energy while the symmetric state is lower. The presence of the BIC at $k_x=0$ can be appreciated from,
    \begin{equation}\label{eq:gamma_pm}
    \text{Im}{[\omega_{\pm}(0)]} \approx -\gamma[1 \pm \cos{(\phi)}].
    \end{equation}    

\subsection{Photonic Jackiw-Rebbi interface state}
Our photonic guided mode Hamiltonian Eqn.~\eqref{eq.phoH} can be transformed into a spinless one-dimensional non-Hermitian Dirac equation with a mass term $m$ through the unitary transformation $\hat{U} = (\hat{\sigma}_x + \hat{\sigma}_z)/\sqrt{2}$,
\begin{equation} \label{eq.Dir}
\hat{U}^\dagger \hat{H} \hat{U} = \hat{H}_\text{Dir} = c \hat{\sigma}_x \hat{p}_x - \hat{\sigma}_z m c^2 - i \gamma.
\end{equation}
Here, $p_x = \hbar k_x$, $c = v$ and $m = \hbar (J \cos{(\phi)}-i \gamma)/v^2$. It has been known for some time in relativistic quantum field theory that the 1D Dirac equation hosts a topologically protected localized state with fractional particle numbers known as a Jackiw-Rebbi mid-gap state ~\cite{Jackiw1976, Niem_PhysRep1986}. Formally, the JR state appears spatially at the interface of two Dirac systems with opposite mass signs $m(x) = m_0 (2H(x) - 1)$  where $H(x)$ is the Heaviside function, and the state wavefunction is written 
\begin{equation}
    |\psi_\text{JR} \rangle = \sqrt{ \frac{c m_0}{2 \hbar}} e^{-c|m(x) \cdot x |/\hbar} \begin{pmatrix} 
    1 \\
    i 
    \end{pmatrix}.
\end{equation}

The JR state is energetically positioned exactly in the center of the gap of the Dirac Hamiltonian. Its topological origin can be appreciated from the fact that no energy mismatch is imposed on the structure. That is, flipping the sign of $m$ does not alter the eigenvalues in Eqn.~\eqref{eq.Dir}, precluding the presence of trivial bound states. A photonic JR state can be constructed if two gratings that differ by $\pi$ in their diffractive coupling phase $\phi$ are put together ~\cite{Lee2021}. The interface between the two gratings corresponds then to a jump $J \to -J$ which can be expressed in terms of the complex mass,
\begin{equation}
    m(x) = \frac{\hbar}{v^2} \left[J (2H(x) - 1) -i \gamma \right].
\end{equation}

The topological origin of the JR state can be traced back to its condensed matter analogue in the electron bands of 1D chains of conjugated polymers, the SSH model ~\cite{Niem_PhysRep1986, heeger1988solitons}. Therein, the topological invariant is known as the quantized Zak phase $\mathcal{Z} = i \int_\text{BZ} \langle \psi_\pm |\partial_k | \psi_\pm \rangle dk \in \{0,\pi\}$ ~\cite{zak1989berry} where $|\psi_\pm \rangle$ are the symmetric and antisymmetric Bloch modes. The integral is over the Brillouin zone and $\langle \psi_\pm |\partial_k | \psi_\pm \rangle$ is integrated over the real space unit cell. In the photonic grating, $\phi = 0 \to \pi$ corresponds to $\mathcal{Z} = 0 \to \pi$ implying that the band inversion represents a topological phase transition which invokes the bulk-boundary correspondence with consequent presence of a midgap JR state. The existence of the JR state and associated Zak phase mismatch can be linked to the surface impedance condition $Z_L + Z_R = 0$, where $Z_{L,R}$ are the surface impedances of the left and right grating at the interface ~\cite{xiao2014surface}. Moreover, the band inversion can also be linked to a different type of a topological invariant in the far-field, given by the half-integer charge of an optical Skyrmion number ~\cite{bouteyre_arxiv2022}. 

\section{Design and fabrication of inverted double-grating structure}
\label{sec:design}

\begin{figure*}
\centering
\includegraphics[width=\linewidth]{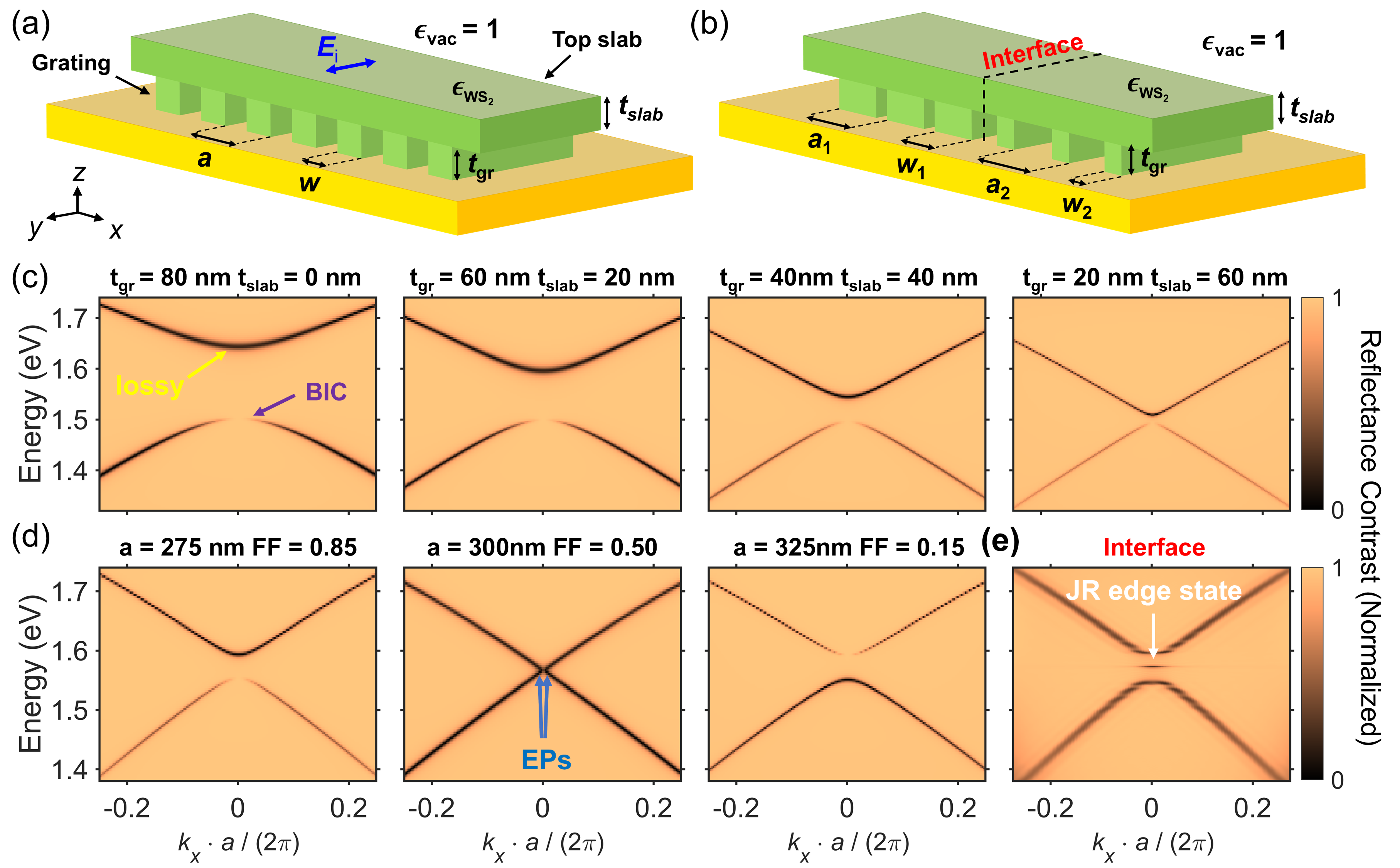}
  \caption{\textbf{Simulated angle-resolved reflectance contrast of \ch{WS_2} single and double inverted grating structures on gold.} \textbf{(a)} Schematic of the single inverted grating structure. $a$, $w$, $t_\mathrm{gr}$, and $t_\mathrm{slab}$ correspond to the grating period, width of \ch{WS2} wires, and thicknesses of the grating and slab respectively. $\epsilon_\mathrm{WS_2}$ and $\epsilon_\mathrm{vac}$ correspond to the \ch{WS2} and vacuum permittivities respectively. Blue double arrow denotes incident polarisation direction $\boldsymbol{E}_\mathrm{i}$ parallel to the grooves of the grating, corresponding to TE excitation. \textbf{(b)} Schematic of the double inverted grating structure with respective periods and widths $a_{1,2}$ and $w_{1,2}$. Interface between the two gratings denoted by the black dashed line. \textbf{(c)} Simulated reflectance contrast of single WS$_2$ grating on gold with increasing thicknesses of $t_\mathrm{slab}$ from left to right leading to reduction of the band gap. Total thickness of the structure is kept constant for each. \textbf{(d)} Simulated tuning of the grating filling factor to achieve photonic band inversion. Period also tuned to shift each mid-gap to the same energy. First three panels correspond to reflectance contrast of single inverted gratings. Left panel shows BIC on the lower energy branch for a high $FF$. Centre-left panel exhibits two exceptional points (EPs) ~\cite{miri2019exceptional} where the modes cross for $FF=0.5$. Centre-right panel shows BIC flipped to the upper energy branch for low $FF$. \textbf{(e)} Simulated reflectance contrast of a double inverted grating structure combining the high and low $FF$ gratings as in \textbf{(b)}. A topologically protected JR interface state is observed at the mid-gap energy as a result.}
  \label{Figure2}
\end{figure*}

Here we explain the rationale behind the unique design of our grating structures, and detail the subsequent fabrication process. As per our analytical model in Section \ref{sec:model}, to obtain photonic band inversion and thus a Jackiw-Rebbi interface state, one needs to tune the grating filling factor in order to close the photonic gap and reopen it. However, the high refractive index of a \ch{WS_2} grating leads to a band gap so large, that no amount of tuning the filling factor can close the gap (see Supplementary Note 2). For this reason, we employ an additional \ch{WS_2} flake on top of our gratings to controllably reduce the effective refractive index contrast between the etched and unetched sections of the gratings, thus reducing the photonic band gap. This design also leverages on the intrinsic vdW adhesive forces of such materials, allowing simple transfer of the top bulk \ch{WS_2} flake (i.e. slab) onto any pre-fabricated nanostructures and devices as illustrated schematically in Figure \ref{Figure2} (a). Here, a gold substrate was chosen owing to its strong reflectivity, thus yielding high contrast with the optical grating modes in reflectance measurements. In addition, gold further illustrates the range of substrates compatible with vdW materials, whilst also acting as a natural etch stop to produce as-designed structures. We note that plasmonic effects are not expected owing to the s-polarised incident light used to excite only TE grating modes.


The effects of adding a top \ch{WS_2} slab to the gratings was simulated using the RCWA technique (see Methods) as in Figure \ref{Figure2} (c). The electric field component of the incident light was fixed along the direction parallel to the grooves (i.e. along $y$ in Figure \ref{Figure2} (a)), corresponding to TE polarisation. Here, the angle-resolved reflectance contrast with respect to the wavevector component $k_x$ from the inverted grating structure is plotted for changing top slab thickness $t_\mathrm{slab}$, whilst keeping the total structure thickness constant. A clear reduction of the band gap is observed with increasing $t_\mathrm{slab}$, corresponding to a decrease in the effective refractive index contrast between the repeating high and low index sections of the structure.

\begin{figure*}
\centering
\includegraphics[width=\linewidth]{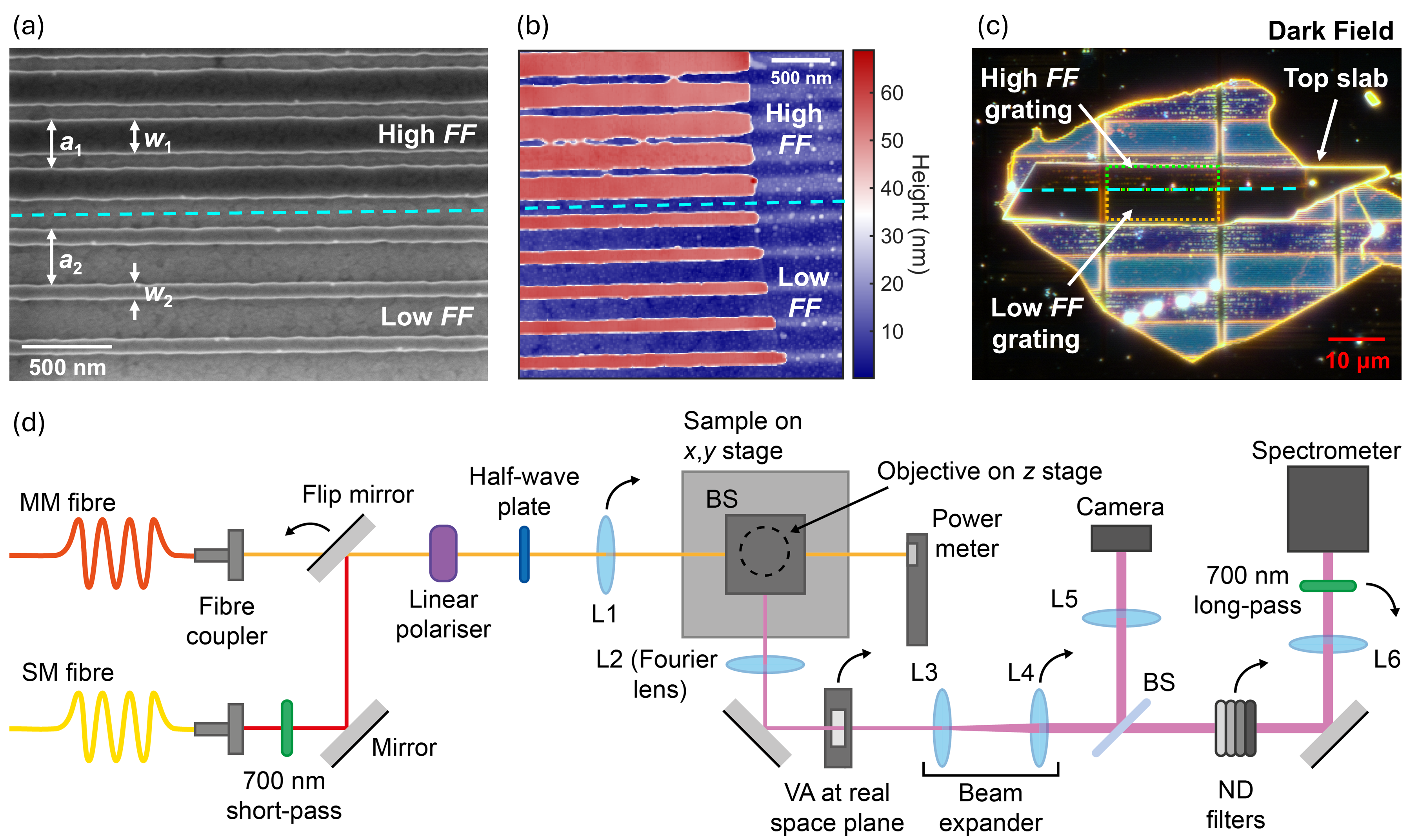}
  \caption{\textbf{\ch{WS_2} double grating structure fabrication and experimental Fourier setup.} \textbf{(a)} SEM image of example double grating interface before top slab transfer. Dashed light blue line corresponds to the interface between high and low FF gratings at the top and bottom of the image respectively. \textbf{(b)} AFM image of double grating interface before top slab transfer. Measured grating thickness $47$ nm. \textbf{(c)} Dark field microscope image of \ch{WS_2} inverted double grating structure with parameters $t_\mathrm{gr}=47$ nm, $t_\mathrm{slab}=41$ nm, $a_{1,2}=279,319$ nm, $F_{1,2}=0.81,0.41$. High and low filling factor inverted gratings denoted by green and orange dotted boxes respectively. \textbf{(d)} Schematic of Fourier spectroscopy setup with angle-resolved reflectance and photoluminescence capability. All optics on flip mounts denoted by curved arrows. Sample sits on the $x,y$ stage beneath the beam splitter (BS) and objective (denoted by black dashed circle) with numerical aperture (NA) $0.7$. For reflectance measurements the multi-mode (MM) fibre is coupled to a white light source with long-pass filter removed. Lens L1 ensures uniform K\"ohler illumination of the sample, whilst L2 images the back focal plane (i.e. Fourier plane) of the objective. Variable aperture (VA) allows selection of signal region from the real-space plane. L3 and L4 act as a beam expander for the Fourier spot. L5 and L6 focus the Fourier plane onto the sample camera and spectrometer slit respectively. For photoluminescence the single-mode (SM) fibre is coupled to a $637$ nm laser with both short- and long-pass filters in the optical path and L1 removed.}
  \label{Figure3}
\end{figure*}

To initiate photonic band inversion, we then simulated a grating with a top slab (i.e. an inverted grating) with parameters $t_\mathrm{gr}=35$ nm and $t_\mathrm{slab}=45$, and varied the filling factor as in Figure \ref{Figure2} (d). The filling factors of the gratings were chosen to obtain three different topological regimes, one with a BIC on the lower energy branch, a completely closed gap with two exceptional points ~\cite{miri2019exceptional}, and one with a BIC on the upper energy branch, as in the left, centre, and right panels respectively. Here, the grating period was also tuned to compensate for the redshift induced by the change of the filling factor, thus ensuring that the mid-gap energy was the same for each structure. Using this inverted grating design, we show that photonic band inversion via closing and re-opening of the band gap is possible. To obtain the JR interface state, we then placed the high and low filling factor gratings from the left and right panels of Figure \ref{Figure2} (d) adjacent to one another, as illustrated by the schematic in Figure \ref{Figure2} (b). Via further reflectance contrast simulations, this time using a Finite-Difference Time-Domain (FDTD) solver (see Methods), a clear state within the photonic band gap was realised as in Figure \ref{Figure2} (e), which we attribute to a topologically-protected JR interface state. 

\begin{figure*}
\centering
  \includegraphics[width=\linewidth]{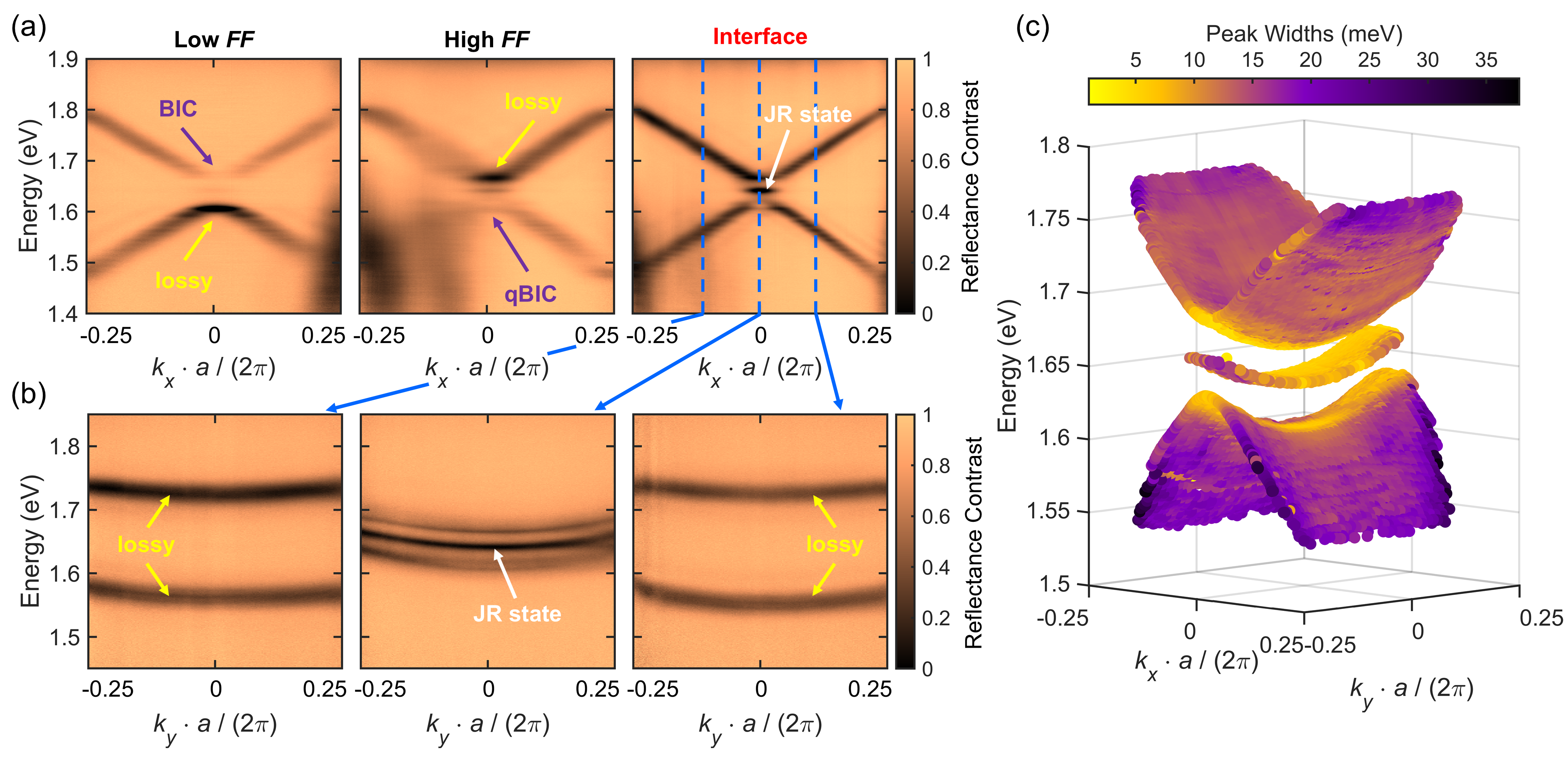}
\caption{\textbf{Experimental reflectance contrast characterisation of the JR interface state.} \textbf{(a)} Angle-resolved reflectance contrast measurements of the low FF grating ($a_2$=330nm, $FF_2$=0.3), high FF grating ($a_1$=261nm, $FF_1$=0.7), and whole structure including the interface respectively, in the $k_x$ direction at $k_y$=0 \textmu m$^{-1}$. Wavevector normalised to the respective grating periods $a$, where the average between the high and low FF gratings is taken for the interface region (right panel). \textbf{(b)} Angle-resolved reflectance contrast of the grating interface region with respect to the $k_y$ direction for three different values of $k_x$, as denoted by the blue dashed lines in the right panel of \textbf{(a)}. \textbf{(c)} Three-dimensional tomographic reconstruction of the grating and interface state modes in momentum space.}
 \label{Figure4}
\end{figure*} 

To realise such inverted double-grating structures in experiment, we began by mechanically exfoliating \ch{WS_2} flakes of a range of thickness ($\sim10-500$ nm) onto electron-beam evaporated gold on silicon wafers (see Methods). Subsequent electron-beam lithography of the substrates coated with a positive resist produced a pattern of alternating high and low filling factor grating pairs. The grating period and electron beam dosage were varied across multiple pairs to account for different flake thicknesses and etching rates. Reactive ion etching using a partially chemical (\ch{SF_6}) and physical (\ch{CHF_3}) etch recipe produced grating structures etched to the gold with parallel wires of \ch{WS_2}. The interface between high and low filling factor gratings were imaged via scanning electron microscopy (SEM) and atomic force microscopy (AFM) as in Figures \ref{Figure3} (a) and (b) respectively, for characterisation of the fabricated period, filling factor, and thickness.

We note that with such thin layers of material used ($<50$ nm), partial etching down to nanometer resolution required to achieve band inversion at the same mid-gap energy would be challenging to realise experimentally. A way around this issue is through our inverted grating design, with both the grating and unetched top flake thicknesses chosen to nanometer precision, thus allowing reliable tuning of the photonic modes.

The transfer step involved first a calibration via simulating the reflectance contrast spectra of the gratings with the measured parameters upon varying top slab thickness. This process enabled accurate determination of the slab thickness required to achieve two topologically distinct adjacent gratings, and thus a JR interface state. \ch{WS_2} was then exfoliated onto PMMA spin-coated on silicon covered with a PVA sacrificial layer. A flake of the required thickness is selected with the help of AFM and transferred onto the gratings, covering the interface (see details in Methods). The resulting inverted double-grating structures on gold were imaged via dark field illumination as shown in Figure \ref{Figure3} (c). Here, the top slab was of thickness $t_\mathrm{slab}=41$ nm, and the grating $t_\mathrm{gr}=47$ nm. The darker blue rectangles of the etched bottom flake correspond to the high filling factor gratings with a period $a_1=279$ nm and a filling factor $F_1=0.81$, and the lighter blue rectangles the lower filling factors gratings with $a_2=319$ nm and $F_2=0.41$. The top \ch{WS_2} slab was positioned to cover at least one of each type of grating, as well as the interface between them for measurement of all three topological regimes. The covered high and low filling factor grating regions (i.e. inverted gratings) of interest are denoted by the green and orange dotted boxes in Figure \ref{Figure3} (c) respectively.

\section{Far-field measurements of Jackiw-Rebbi interface state}
\label{sec:farfield}

Angle-resolved reflectivity contrast measurements of the structure were obtained using a Fourier spectroscopy setup as depicted schematically in Figure \ref{Figure3} (d). A white light illumination source was used with TE configuration to compare to the simulated spectra (more details in Methods). The reflectivity contrast maps in the $k_x$ direction (for $k_y=0$) of the low FF grating region, interface, and high FF grating region are presented in Figure \ref{Figure4} (a). In agreement with our previous simulations from Figure \ref{Figure2} (d), one can observe a BIC on the high energy branch of the spectra from the left panel of Figure \ref{Figure4} (a) corresponding to the low FF grating, and a (quasi-) BIC flipped to the low energy branch for the high FF grating in the central panel of Figure \ref{Figure4} (a). Here we observe a quasi-BIC at $k_x=0$ as the top \ch{WS2} slab did not fully cover the entire high FF grating (green dotted box in Figure \ref{Figure3} (c)), thus leading to a finite linewidth and obscured reflectance at negative $k_x$. Additional experimental data from a separate structure are presented in Supplementary Note 3 exhibiting true BICs for both the high and low FF gratings, whilst also highlighting the repeatability of our transfer fabrication method, and tunability of the bands in energy. By measuring at the interface region shown in the right panel of Figure \ref{Figure4} (a), there is a clear feature at the mid-gap energy, not present in the two gratings alone. These experimental results therefore not only confirm the successful photonic band inversion via tuning of the filling factor, but also the different topological phases of such inverted band structures, resulting in the formation of an interface-localised Jackiw-Rebbi state within the band gap.

We further considered the reflectance contrast along the $k_y$ direction by rotating the sample whilst keeping the TE excitation configuration as in Figure \ref{Figure4} (b). Three measurements were taken at the interface region for three different values of $k_x$ as denoted by the dashed blue lines in the right panel of Figure \ref{Figure4} (a). For all three sets of reflectance contrast spectra, we observe that the dispersion of the two lossy grating modes is parabolic with positive curvature. As $k_x$ is reduced to zero, the JR state emerges and follows the parabolic dispersions of the two lossy modes, remaining at the mid-gap energy with positive curvature in the $k_y$ direction. Via fitting to a Lorentzian function we extract a linewidth of $10$ meV, and further calculate that the JR state yields an angular emission bandwidth in the direction perpendicular to the grooves of just $\Delta\theta_x=8.0\degree$, yet can be in- and out-coupled throughout the full measurement region in the direction parallel to the grooves of $\Delta\theta_y=88\degree$. These results therefore highlight the highly confined nature of the JR interface state in TMD-based inverted double-gratings, with strong localisation in both energy and momentum space.

We subsequently plot the full 3D tomographic reconstruction of the experimental grating mode structure in $k_{x,y}$ as in Figure \ref{Figure4} (c). One can observe that the lower energy mode has a saddle shape, with negative curvature in $k_x$ but positive curvature in $k_y$. In contrast, the higher energy mode displays a 3D parabolic shape with positive curvature in both directions, but greater curvature in $k_x$. In addition, the JR state is a parabolic stripe along $k_y$, with a width of $1.1$ \textmu m$^{-1}$ in $k_x$. As well as being able to visualise the full dispersion of all modes in momentum space, this three dimensional reconstruction also highlights the relative linewidths of each mode in energy. We expect the BICs at the band edges and JR state to be most strongly confined to the structure for $k_x=0$, owing to the symmetry conditions required to host such states ~\cite{van2021unveiling,gao2022dark}. This can be observed experimentally via the narrower linewidths of the modes along the line $k_x=0$, but also shows that even at large angles in the direction parallel to the grooves up to $\pm44\degree$ from the normal, the linewidths of the resulting modes changes negligibly. Such 1D double grating structures therefore possess highly selective directivity profiles depending on the plane of incidence.

\section{Near-field measurements of Jackiw-Rebbi interface state}
\label{sec:nearfield}

\begin{figure*}
\centering
\includegraphics[width=\linewidth]{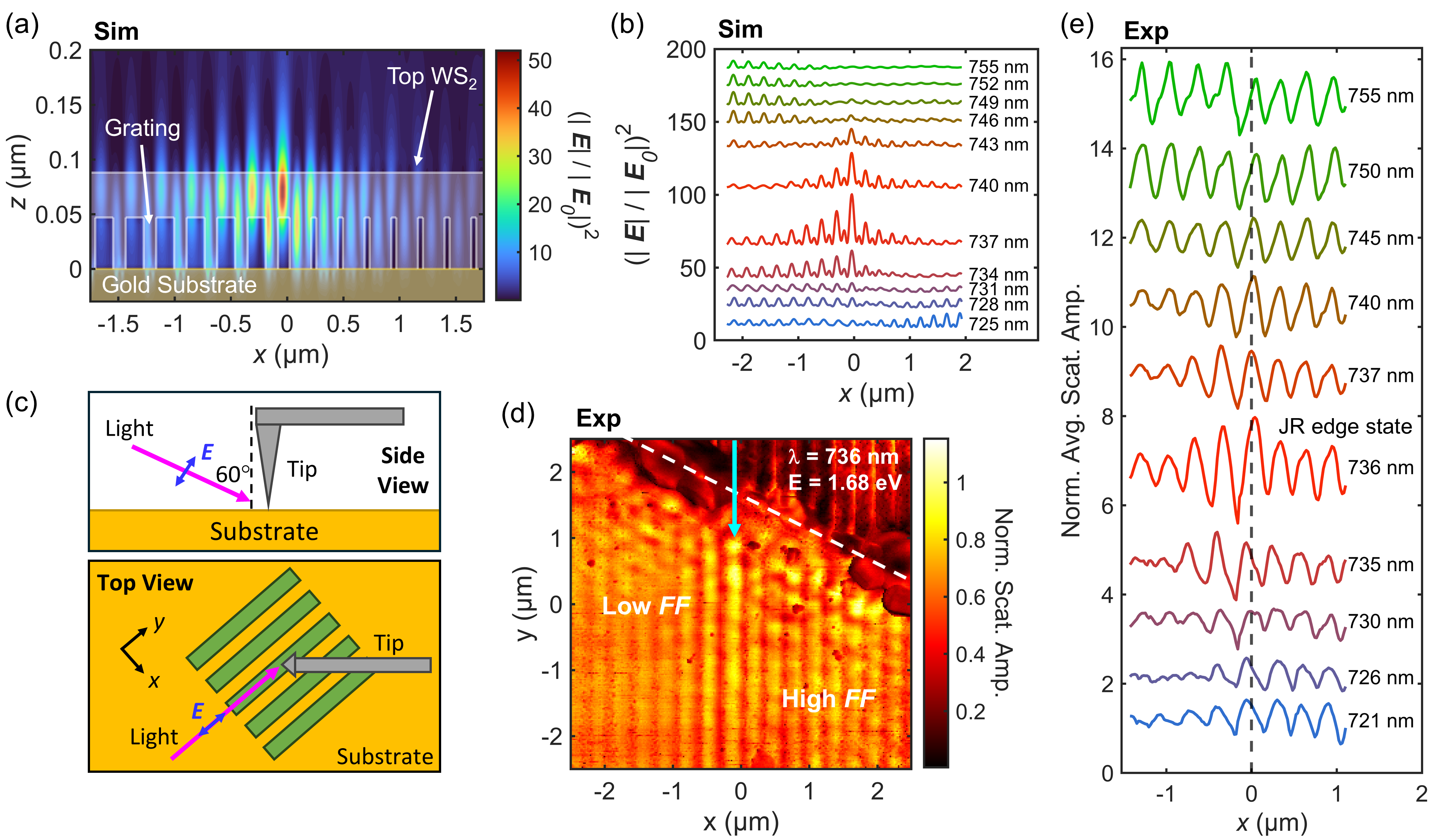}
  \caption{\textbf{Near-field study of the JR interface state.} \textbf{(a)} FDTD simulation of the near-field distribution of the JR state for 60\degree incident illumination angle. \textbf{(b)} Profiles of the simulated electric field strength along the top surface of the structure for changing incident wavelength. \textbf{(c)} Schematic of the s-SNOM setup from side (top panel) and top-down (bottom panel) views. Incident light depicted by the pink arrow propagating at $60\degree$ to the tip axis. Blue double arrow denotes polarisation direction (p-polarised) with component along the tip axis. Grating oriented parallel to the incident light to excite along the $k_y$ direction. \textbf{(d)} Experimental normalised near-field scattering amplitude image at $736$ nm excitation of the \ch{WS_2} double grating structure. Light blue arrow highlights interface between low and high FF gratings where near-field enhancement is observed. Dashed white light corresponds to edge of the top \ch{WS_2} slab. \textbf{(e)} Averaged near-field scattering amplitude profiles over multiple rows in the y direction for varying incident wavelength. Grating interface centered at $x = 0$ showing enhancement only around $736$ nm illumination.}
  \label{Figure5}
\end{figure*}

From our RCWA simulations and experimental angle-resolved reflectivity measurements, the existence of a topological photonic JR interface state in \ch{WS_2} double inverted grating structures is clearly demonstrated in the far-field. We now investigate the near-field localization of this state through both simulation, and experimental probing of the local electric field distribution via s-SNOM. 

Figure \ref{Figure5} (a) depicts the simulated electric field confinement in a cross-sectional slice through the middle of the double gratings as calculated via FDTD. To emulate an s-SNOM measurement, the incident excitation was a p-polarized plane wave travelling at $60\degree$ to the substrate normal, in the direction parallel to the grooves (see Methods). It is clear that the JR state is strongly localized at the interface between the two topologically distinct gratings, with an electric field enhancement of up to $50$ times that of the incident wave. In addition, we observe that confinement is highest within the top WS$_2$ slab rather than the grating itself, which we attribute to the higher overall effective refractive index of the slab portion. The field also protrudes significantly out of the top of the structure, enabling direct probing via a nanoscale tip.

We then simulated a range of illumination wavelengths, and plot profiles of the electric field strength along the top surface of the structure, i.e. where we are able to probe with an s-SNOM tip, as in Figure \ref{Figure5} (b). These simulations show that the JR state peaks in electric field confinement at around $737$ nm ($1.68$ eV). We also note that at lower wavelengths, the overall field strength is higher to the right of the grating interface ($x = 0$) which corresponds to the lossy mode of the high FF grating, whereas the fields are weaker to the left of the interface corresponding to the BIC. This behaviour flips as we tune through the JR state to higher wavelengths owing to the band inversion between the high and low filling factor gratings.

\begin{figure*}[!ht]
\centering
\includegraphics[width=\linewidth]{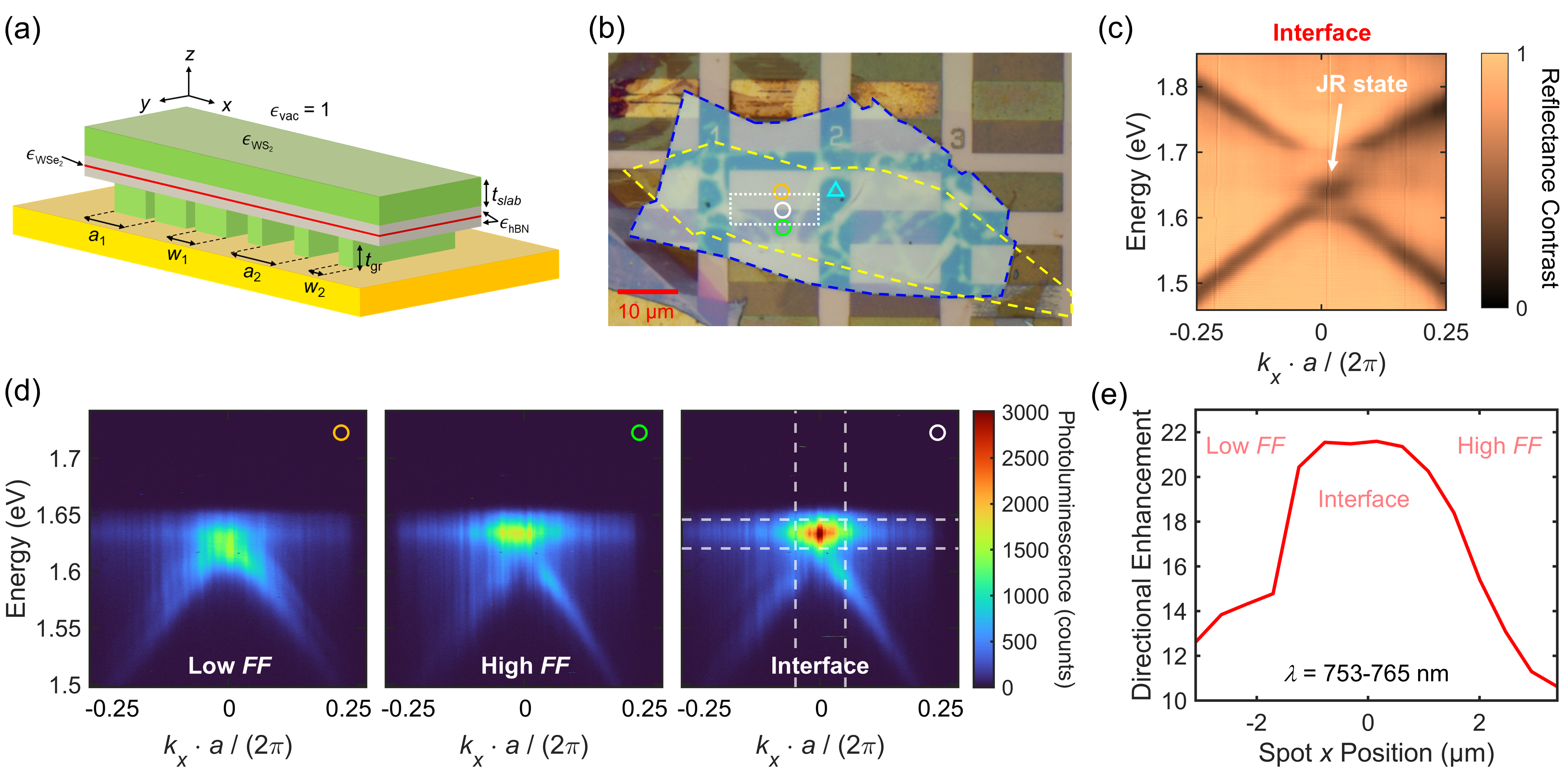}
  \caption{\textbf{PL enhancement via active grating heterostructure} \textbf{(a)} Schematic of the active double grating heterostructure composed of hBN-encapsulated \ch{WSe_2} between a \ch{WS_2} double grating and top slab on a gold substrate. \textbf{(b)} Optical microscope image of the fabricated heterostructure. Top \ch{WS_2} slab and monolayer \ch{WSe_2} outlined by dashed blue and yellow lines respectively. White dashed box corresponds to Fourier RC measurement region. Coloured circles correspond to the Fourier PL excitation spots. Light blue triangle denotes the Fourier PL reference measurement location on encapsulated monolayer \ch{WSe_2} away from the gratings. \textbf{(c)} Angle-resolved RC taken from the white dashed region in \textbf{(b)} around the double grating interface showing JR state within the photonic band gap. \textbf{(d)} Angle-resolved photoluminescence measurement of the low FF grating, high FF grating and the interface between the two corresponding to the orange, green, and white circles from \textbf{(b)} respectively. \textbf{(e)} Photoluminescence directional enhancement factor along the direction perpendicular to the grooves compared to uncoupled monolayer. Signal integrated over $k_x$ and energy as shown by the dashed white lines in \textbf{(d)}.}
  \label{Figure6}
\end{figure*}

To successfully probe the JR interface state via experimental s-SNOM measurements, the sample had to be precisely orientated owing to its strong directivity. With the high incident illumination angle to the substrate normal (see s-SNOM Methods), excitation of the JR state was only possible via light propagating in the plane of incidence parallel to the grooves of the grating, as depicted by the pink arrow in Figure \ref{Figure5} (c). This illumination angle corresponds to zero wavevector in the $k_x$ direction, with mode dispersion along $k_y$ given by the central panel of Figure \ref{Figure4} (b). We further extrapolated the JR state dispersion in $k_y$ up to $60\degree$ incident angle (see Supplementary Note 4), yielding an estimated excitation energy of $1.68$ eV ($735\text{-}738$ nm), which agrees perfectly with our simulations from Figure \ref{Figure5} (b). Importantly, this excitation regime means a component of the incident electric field lies parallel to the grooves as shown by the blue double arrow in Figure \ref{Figure5} (c), which is required to excite the TE-polarised grating modes and thus the JR interface state.

Following careful sample alignment, we performed s-SNOM measurement at $736$ nm excitation as in Figure \ref{Figure5} (d), and observed enhanced scattering from the interface between the two gratings as marked by the light blue arrow. The data shown have been median line levelled separately either side of the grating interface to minimise the intrinsic material response, which differs depending on the effective refractive index experienced by the tip. We thus focus on scattering arising mostly from probing the localized electric fields associated with the grating modes, as described in more detail in Supplementary Note 5.

We then repeated the scan for a range of wavelengths and plotted line profiles of the scattering intensity averaged over multiple rows in the $y$ direction as in Figure \ref{Figure5} (e), where the grating interface and thus JR state is centred at $x = 0$. We observe that the peak in scattering at $736$ nm along the dashed line quickly decays when changing the incident wavelength, which correlates very well with our electric field profile simulations. The localization of the enhanced scattering exactly where we expect it spatially, also at the correct energy we predicted from our far-field measurements for light incident at 60\degree, strongly suggests that we have successfully probed the near-field response of this topological photonic JR interface state, confirming its highly localized nature in real space, $k$-space, and energy.

\section{Photoluminescence enhancement at the Jackiw-Rebbi state}
\label{sec:pl}

After demonstrating the near-field localization of the JR interface state in both simulation and experiment, we now study the PL enhancement and directivity of an emitter coupled to such a state. To do so, we consider a similar double grating structure shown schematically in Figure \ref{Figure6} (a), in which an hBN-encapsulated \ch{WSe_2} monolayer is embedded between the grating layer and the top \ch{WS_2} slab. The thin hBN layers prevent charge transfer between the \ch{WSe_2} monolayer and bulk \ch{WS_2} with negligible effect on the grating mode structure, thus enabling bright photoluminescence emission. A combination of PDMS and PMMA-based transfer techniques was used to build this structure as explained in more detail in Methods. Figure \ref{Figure6} (b) presents an optical microscopy image of the fabricated double-grating heterostructure, in which the top \ch{WS_2} flake and encapsulated \ch{WSe_2} monolayer are indicated by the blue and yellow dashed lines respectively. Here we used a top flake of thickness $t_\mathrm{slab}=32$ nm with grating $t_\mathrm{gr}=69$ nm. The top and bottom hBN layers were of $3$ and $6$ nm thicknesses respectively. To confirm the presence of a JR interface state in this more complex heterostructure, we performed angle-resolved reflectance contrast measurements as with the previous sample. Figure \ref{Figure6} (c) shows results from the grating interface, taken from the white dashed rectangular region of Figure \ref{Figure6} (b) using the variable aperture. A clear JR state is once again visible within the photonic band gap, shifted slightly to lower energy likely owing to coupling with the \ch{WSe_2} exciton which exists at $1.65$ eV.

Photoluminescence measurements were performed at room temperature using the Fourier setup from Figure \ref{Figure3} (d), with the introduction of a $637$ nm laser for excitation. The measurements were taken from three different locations indicated by the circles in Figure \ref{Figure6} (b), corresponding to the two separate low and high filling factor gratings, and the grating interface. Clear coupling of the \ch{WSe_2} excitonic emission to the lossy grating mode can be seen in the left panel of Figure \ref{Figure6} (d) for the low FF grating, where enhanced signal follows the dispersion of the modes as measured via angle-resolved reflectance contrast in Figure \ref{Figure6} (c). In the centre panel a dip in photoluminescence can be observed around $k_x=0$ at $\sim1.6$ eV, corresponding to emission unable to outcouple to the far-field from the BIC within the high FF grating. Some enhancement is also observed around $1.64$ eV, however this corresponds to leaked signal from the JR state owing to the close spatial proximity of the measurement regions, and the finite extent of the JR state electric field profile. Finally, the right panel of Figure \ref{Figure6} (d) shows the strongest PL enhancement from the grating interface region at the JR state energy. The PL is enhanced over the same range of $k_x$ as the JR state in reflectance contrast, therefore indicating coupling of the \ch{WSe_2} excitonic emission to the state.

To quantify the directional photoluminescence enhancement from monolayer \ch{WSe2} coupled to the JR state, the PL intensity was integrated over the region depicted by the dashed white lines in the right panel of Figure \ref{Figure6} (d). The size of this integration region is equal to the fitted linewidth of the JR state-coupled PL peak in both $k_x$ and energy. This integration was repeated for PL dispersions measured at multiple $x$ positions (vertical direction in Figure \ref{Figure6} (b)), and then divided by the integrated counts from a PL measurement taken at a reference position as marked by the light blue triangle in Figure \ref{Figure6} (b). In this region, the bottom \ch{WS_2} remains unetched, resulting in an hBN-encapsulated monolayer \ch{WSe_2} embedded between two \ch{WS_2} slabs (as detailed further in Supplementary Note 6), which are transparent at the \ch{WSe2} emission energy of $1.65$ eV. The reference PL emission was therefore not coupled to any of the grating or interface modes, with no dispersion in $k$-space (see Supplementary Note 6). As such, the directional enhancement factor corresponds to the ratio of PL from the grating heterostructure between $753-765$ nm wavelength emitted into a wavevector range of $\pm1.1$ \textmu m$^{-1}$ (angular bandwidth of $\pm7.5\degree$), to that of the PL from the reference region over the same wavelength and wavevector range. Note that the wider angular bandwidth for the active grating heterostructure compared to the passive device presented in Figure \ref{Figure4} (a) is likely owing to losses induced by the monolayer \ch{WSe2}, and the presence of hBN which reduces the overall effective refractive index and thus broadens the modes. The directional enhancement results are plotted in Figure \ref{Figure6} (e), and give an indication of how much additional emission is directed upwards of the sample plane via JR state coupling compared to the omnidirectional emission from uncoupled monolayer \ch{WSe2}. We calculate a directional enhancement factor of between $11$ and $13$ for monolayer coupled to the separate gratings modes, and up to $22$ for the JR state-coupled emission at the grating interface.

%
%
%
%

\section{Conclusion}
\label{sec:conclusion}

In conclusion, we realized here precise tuning and band inversion of 1D photonic grating modes in stacked van der Waals devices, yielding topologically-protected Jackiw-Rebbi interface states in both simulation and experiment with strong theoretical agreement. Via far-field angle-resolved reflectance spectroscopy, we fully characterised the grating mode dispersions throughout $k_{x,y}$, revealing strongly directional emission of the JR state along the direction perpendicular to the grooves, with an angular bandwidth of $8.0\degree$ and linewidth $10$ meV. Such JR states also exhibited clear spatial confinement to the boundary between topologically distinct gratings as detected using scattering-type scanning near-field optical microcopy. We further fabricated an active five-layer double grating heterostructure incorporating hBN-encapsulated monolayer \ch{WSe_2}, demonstrating directional enhancement of the exciton photoluminescence via coupling to the JR state. Our results thus highlight the utility of vdW materials for easily integrable and adaptable multi-layer on-chip devices, able to enhance and shape the radiation patterns of coupled emitters, with in-built topological protection.

\section*{Acknowledgements}
\label{sec:acknowledgements}
SR, AIT, PB, XH and AK thank EPSRC grants EP/V006975/1, EP/V007696/1, EP/V026496/1, EP/S030751/1. Yue Wang thanks EPSRC grant EP/V047663/1 and Royal Academy of Engineering fellowship RF/201718/17131.
HS acknowledges the Icelandic Research Fund (Rannís), grant No. 239552-051 and the project No. 2022/45/P/ST3/00467 co-funded by the Polish National Science Centre and the European Union Framework Programme for Research and Innovation Horizon 2020 under the Marie Skłodowska-Curie grant agreement No. 945339.
KW and TT acknowledge support from the JSPS KAKENHI (Grant Numbers 21H05233 and 23H02052), the CREST (JPMJCR24A5), JST and World Premier International Research Center Initiative (WPI), MEXT, Japan.
RG acknowledge support of ERC Consolidator grant QTWIST (no. 101001515), European Quantum Flagship Project 2DSIPC (no. 820378) and EPSRC (grant numbers EP/V007033/1 and EP/Z531121/1).

\section*{Methods}
\label{sec:methods}
\textbf{Simulations.} Angle-resolved reflectance simulations of single gratings were performed using an open source Python implementation of the Rigorous Coupled-Wave Analysis (RCWA) technique named $S^4$ ~\cite{Liu2012s4}. A unit cell was defined with periodicity of the grating, composed of stacked layers for the substrate, grating, top slab, and superstrate. Experimentally measured \ch{WS2} anisotropic refractive index data was used from Ref. ~\cite{munkhbat2022optical}. The grating layer was imprinted with a pattern according to the desired filling factor to achieve a periodically modulated structure. TE polarisation was considered for the excitation source, with incident angle varied in the direction perpendicular to the grooves (i.e. $x$). Reflected light was subsequently normalised to a reference simulation of just the substrate to achieve reflectance contrast.

Double gratings were simulated using the Finite-Difference Time-Domain (FDTD) method through the commercial software package Lumerical. $20$ periods of each adjacent grating were simulated to ensure accuracy, with the simulation region encompassing a 2D slice through the cross-section of the structure. Plane wave source with TE polarisation was swept over a range of incident angles and reflected light intensity measured, as with the RCWA simulations. Perfectly-matched layer (PML) boundaries were used along all directions to reduce any reflections at the simulation edges. Electric field profiles taken using frequency-domain field and power monitors.

\textbf{Fabrication.} \ch{WS2} flakes from HQ graphene were mechanically exfoliated onto electron-beam evaporated $150$ nm gold + $10$ nm Ti on Si substrates. $105\degree$C temperature used to promote adhesion, before cooling and then exfoliating. A positive resist (ARP-9 AllResist
GmbH) was then spun onto the substrates at $3500$ rpm for $60$ s, before heating for
$2$ minutes at $180\degree$C. Electron beam lithography was then performed using a Raith GmbH
Voyager system at $50$ kV accelerating voltage and $560$ pA beam current, producing patterns of alternating high and low filling factor grating pairs. Multiple pairs with varying electron beam dosage and period were patterned onto each flake to account for different thicknesses and variations in etch rates. Reactive ion etching for $40$ s with $0.14$ mbar pressure and a DC bias of 135 V was used with a partially chemical and physical etch recipe of both \ch{SF6} and \ch{CHF3} gases. This produced parallel wires of \ch{WS2} without a preferential zig-zag etch, as with purely chemical etching.
Subsequent layers of multi-layer hBN (NIMS) and \ch{WS2} for transfer were exfoliated onto PMMA-coated Si substrates with a PVA sacrificial layer between. Flakes of ideal thickness were identified via atomic force microscopy, before scribing of a circle $\sim3$ mm diameter around the desired flake. The PVA was then dissolved with deionised water, and the floating membrane 'fished' with a metal ringed cantilever. Inside a nitrogen-environment glovebox, the membrane was then brought into contact with the desired grating via a 3D stage with vacuum arm, and heated to $180\degree$C to melt the PMMA. Acetone/IPA washing followed by a brief \ch{O2}-plasma cleaning removed the remaining membrane, leaving flakes adhered to the gratings.
Monolayer \ch{WSe2} (HQ Graphene) was exfoliated onto PDMS stamps and transferred using the same glovebox transfer setup. $65\degree$C temperature was used on place-down of the stamp to promote adhesion. Reduction of the heat upon lift-off of the PDMS caused detachment of the \ch{WSe2} from the stamp, and adherence to the target grating heterostructure sample.

\textbf{Angle-Resolved Reflectance.} A home-built Fourier spectroscopy setup was used for experimental angle-resolved measurements of the grating structures as depicted in Figure \ref{Figure3} (d). White light illumination was used with polarisation always along the grooves of each grating (TE) dictated by a linear polariser and half-wave plate combination. A 50:50 beam splitter subsequently redirected light to an objective lens with numerical aperture $0.7$. A lens placed at its focal length from the back focal plane of the objective was used to image the Fourier plane. A variable rectangular aperture was placed at the real-space plane after the first lens to selectively collect signal from specific spatial regions of each sample. A further two-lens telescope was used to enlarge the Fourier image, before focusing onto the spectrometer slit by a final lens. Closing of the spectrometer slit to $50$ \textmu m allowed projection of a single slice in Fourier space onto the charge-coupled device (CCD), at a single $k_y$ for a range of $k_x$, or vice versa by rotating the sample $90\degree$. As with the simulations, each grating reflectance measurement was normalised to the signal from gold substrate alone to yield reflectance contrast.

\textbf{Angle-Resolved Photoluminescence.} The same Fourier setup was used for photoluminescence measurements by changing the excitation source to a $637$ nm Vortran Stradus diode laser. Additional short- and long-pass filters were flipped into the beam path before and after the sample respectively, so that only the angle-resolved PL signal was collected by the spectrometer.

\textbf{Near-Field Imaging.} 
s-SNOM data were collected under ambient conditions utilising a neaSCOPE microscope from Attocube Systems AG/Neaspec. The neaSCOPE system was coupled with a Chameleon Compact OPO-Vis system from Coherent, consisting of a Ti:Sapphire laser and optical parametric oscillator (OPO) with a tuneable, single wavelength output selectable anywhere between 360-1600nm. Light from the laser was sent into a Michelson interferometer housed within the neaSCOPE system, with an AFM in one arm of the interferometer operating under tapping mode and a clean reference mirror housed in the other. The laser light was focused onto the tip of a metal-coated AFM cantilever (ARROW-EFM from Apex Probes, PtIr coating, tapping frequency 82-85 kHz, tapping amplitude 34-37 nm, tip radius of curvature $\sim25$ nm) at an angle of $60\degree$ to the tip axis via a parabolic mirror. The incident polarisation was aligned along the tip axis, i.e. p-polarised, thus exciting surface plasmon-polaritons at the metal/air boundary which were then used to probe the sample as it was raster scanned beneath. Further laser light scattered off the tip-sample interaction region was collected back into the interferometer before interfering with the light reflected off the reference mirror at the detector. The tapping motion of the AFM cantilever allowed the optical signal from the tip-sample interaction region to be demodulated at harmonics of the tapping frequency (order $3$), thereby reducing the influence of background scattering. Compete background removal was obtained by oscillating the reference mirror at a fixed frequency, resulting in frequency mixing and the formation of side bands in the optical power spectrum that were then tracked.

\section*{Supplemental Documents}
\label{sec:si}

The following files are available free of charge: derivation of full photonic Dirac Hamiltonian and associated eigenvectors; angle-resolved reflectance contrast simulations of \ch{WS2} gratings on gold without top \ch{WS2} slab; experimental angle-resolved reflectance contrast measurements of JR state in an additional fabricated double inverted grating structure; extrapolated experimental mode dispersion of JR state in $k_y$ direction; description of s-SNOM probing regimes; double grating heterostructure experimental PL directional enhancement analysis methodology.

\bibliography{main}

\end{document}


\title{Supplementary Information for: Topological Jackiw-Rebbi States in Photonic Van der Waals Heterostructures}

\author{Sam A. Randerson}
\email{s.a.randerson@sheffield.ac.uk}
\affiliation{School of Mathematical and Physical Sciences, The University of Sheffield, Sheffield S3 7RH, U.K.} 
\author{Paul Bouteyre}
\thanks{S.A. Randerson and P. Bouteyre contributed equally to this work as first authors.}
\affiliation{School of Mathematical and Physical Sciences, The University of Sheffield, Sheffield S3 7RH, U.K.}
\author{Xuerong Hu}
\affiliation{School of Mathematical and Physical Sciences, The University of Sheffield, Sheffield S3 7RH, U.K.}
\author{Oscar J. Palma Chaundler}
\affiliation{School of Mathematical and Physical Sciences, The University of Sheffield, Sheffield S3 7RH, U.K.} 
\author{Alexander J. Knight}
\affiliation{School of Mathematical and Physical Sciences, The University of Sheffield, Sheffield S3 7RH, U.K.}
\author{Helgi Sigur{\dh}sson}
\affiliation{Science Institute, University of Iceland, Dunhagi 3, IS-107 Reykjavik, Iceland}
\affiliation{Institute of Experimental Physics, Faculty of Physics, University of Warsaw, ul. Pasteura 5, PL-02-093 Warsaw, Poland}
\author{Casey K. Cheung}
\affiliation{Department of Physics and Astronomy, The University of Manchester, Oxford Road, Manchester, M13 9PL, U.K.}
\affiliation{National Graphene Institute, The University of Manchester, Oxford Road, Manchester, M13 9PL, U.K.}
\author{Yue Wang}
\affiliation{School of Physics Engineering and Technology, University of York, York YO10 5DD, U.K.}
\author{Kenji Watanabe}
\affiliation{Research Center for Electronic and Optical Materials, National Institute for Materials Science, 1-1 Namiki, Tsukuba, 305-0044 Japan}
\author{Takashi Taniguchi}
\affiliation{Research Center for Materials Nanoarchitectonics, National Institute for Materials Science, 1-1 Namiki, Tsukuba, 305-0044 Japan}
\author{Roman Gorbachev}
\affiliation{Department of Physics and Astronomy, The University of Manchester, Oxford Road, Manchester, M13 9PL, U.K.}
\affiliation{National Graphene Institute, The University of Manchester, Oxford Road, Manchester, M13 9PL, U.K.}
\author{Alexander I. Tartakovskii}
\email{a.tartakovskii@sheffield.ac.uk}
\affiliation{School of Mathematical and Physical Sciences, The University of Sheffield, Sheffield S3 7RH, U.K.} 

\date{\today}

\pacs{}

\maketitle

\newpage
\section{Supplementary Note 1}
\label{sec:SI1}
A more general form of the photonic Dirac Hamiltonian given in Eq.~(1) in the main text reads~\cite{Sigurdsson_Nanopho2024}, 
\begin{equation} \label{eq.phoH}
    \hat{H} = \begin{pmatrix}
        v k_x & Je^{i \phi_2} \\ 
        Je^{-i \phi_2} & -vk_x
    \end{pmatrix}
    - i \gamma \begin{pmatrix}
        1 & e^{i 2\phi_1} \\
        e^{-i 2\phi_1} & 1
    \end{pmatrix}
\end{equation}
Here, we have explicitly written out the phases $\phi_{1,2}$ belonging to first- and second-order diffractive coupling channels which are given by their corresponding Fourier coefficients over the periodic potential $J e^{i \phi_2} = U_2 =  \langle e^{i(k_x + K) x} | u(x) | e^{-i(k_x-K)x} \rangle$ and $\gamma \propto |U_1|^2$ and $\phi_1 = \text{arg}{(U_1)}$ where $U_1 = \langle e^{ik_x x} | u(x) | e^{-i(k_x-K)x} \rangle$. Physically, the phases represent interference between counterpropagating modes ($\phi_2$) and propagating modes with the lossy Fabry-Perot modes around normal incidence ($\phi_1$). The second anti-Hermitian term in $\hat{H}$ is derived by adiabatically eliminating the dynamics of the lossy Fabry-Perot mode by assuming it decays much faster decay compared to the guided modes  $| e^{-iqx} \rangle$ which are protected by total internal reflection~\cite{Sigurdsson_Nanopho2024}. The effective loss rate of the folded guided modes across the $\Gamma$-point is written $\gamma = |c_0|^2 |U_1|^2/\gamma_0$ where $c_0 = \langle \chi^{(0)}(z) | \chi_K(z) \rangle$ describes the vertical overlap (confinement integrals) between the lossy mode $| \chi^{(0)}(z) \rangle$ and the guided mode $| \chi_K(z) \rangle$.

The mirror symmetry of the grating $u(x) = u(-x)$ implies that the Fourier coefficients are real-valued. This means that $\phi_{1,2} \in \{0,\pi\}$ and therefore $e^{2i \phi_1} = 1$ and $e^{i \phi_2} = \pm 1$. This can be understood from the Fourier expansion of the periodic potential
\begin{equation}
    u(x) = \sum_n U_n e^{i K_n x}, \qquad K_n =  \frac{2 \pi n}{a}.
\end{equation}
where the coefficients $U_n$ belong to the Fourier transform,.
\begin{equation}
    U_n = \frac{1}{a} \int_{-a/2}^{a/2} u(x) e^{iK_n x} \, dx.
\end{equation}
Because $u(x) = u(-x)$ is an even function, only the cosine from the $e^{iK_n x}$ contributes which means $U_n \in \mathbb{R}$. The sign of the Fourier coefficients $U_n$ depends on the period of the potential $u(x)$ which means that the phases $\phi_{1,2} \in \{0,\pi\}$ of the diffractive and lossy channel coupling can be flipped from $0 \to \pi$ by adjusting the pitch and filling factor of the grating \cite{Lu_PhotRes2020, Lee2021, Sigurdsson_Nanopho2024}. Importantly, we always have that $e^{2i \phi_1} = 1$ which means that the presence of symmetry protected photonic bound state in the continuum (BIC) is guaranteed at the $\Gamma$-point in the antisymmetric energy branch \cite{Azzam_AdvOptMat2021} topologically tied to a polarization vortex in the far field \cite{Doeleman2018}.

The eigenvectors of $\hat{H}$ for $\phi = \phi_2$ and $e^{2i \phi_1} = 1$ (coinciding with our equation in the the main text) are written, 
\begin{equation}
    \boldsymbol{v}_{\pm}(k_x) = \frac{\pm 1}{A_\pm} \begin{pmatrix}
        h_{\pm} \\ J - i \gamma 
    \end{pmatrix}
\end{equation}
where $h_\pm = vk_x \pm \sqrt{(vk_x)^2 + J^2 - \gamma^2 - 2 i J \gamma \cos{(\phi)}}$ and $A_\pm > 0$ are normalization constants. Specifically, for $\phi=0$ and $k_x=0$ we have, 
\begin{equation}
    \boldsymbol{v}_{\pm}(0) = \frac{J - i \gamma}{\sqrt{2(J^2 + \gamma^2)}} \begin{pmatrix}
        \pm 1 \\ 1 
    \end{pmatrix}
\end{equation}
The $\pm$ here refers to a symmetric and antisymmetric standing-wave of the forward $| e^{i K x} \rangle = (1,0)^\text{T}$ and backward $| e^{-i K x} \rangle = (0,1)^\text{T}$ counter-propagating photons.

\newpage
\section{Supplementary Note 2}
\label{sec:SI2}

\begin{figure*}[ht!]
\centering
  \includegraphics[width=\linewidth]{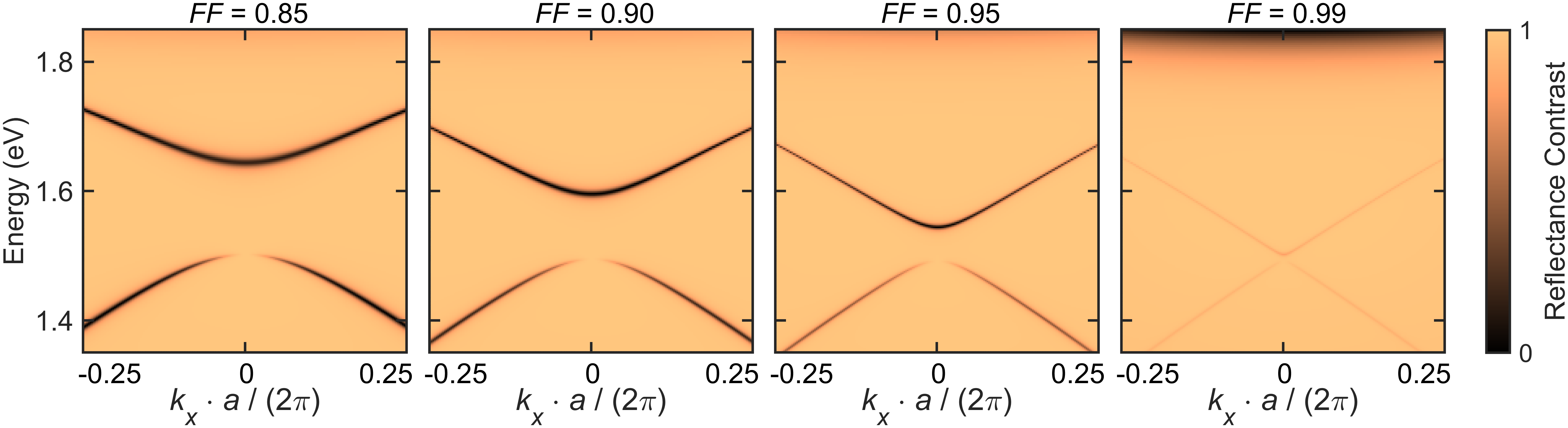}
\caption{\textbf{Simulated reflectance contrast of \ch{WS2} grating on gold with changing filling factor.} From left to right: $FF=0.85,0.90,0.95,0.99$. Grating period and thickness constant for each panel with $a=300$ nm and $t_{gr}=80$ nm respectively. Band gap closes as FF is increased but can never be closed.}
 \label{fig:SI2}
\end{figure*}

We demonstrate here the inability to fully close the associated photonic band gap using high index \ch{WS2} gratings on gold. Figure \ref{fig:SI2} presents Rigorous Coupled-Wave Analysis (RCWA) simulations of the angle-resolved reflectance contrast from grating structures with increasing filling factor (FF). Since the refractive index of \ch{WS2} is large ($\sim4.15$ \cite{munkhbat2022optical}), the index contrast between the high and low index sections of the grating is significant, leading to a wide band gap. Increasing the filling factor in simulation tunes the coupling between the upper and lower photonic modes, thus reducing the band gap. However, as seen from Figure \ref{fig:SI2}, the band gap for such structures can never fully close, no matter how high the filling factor is tuned. Besides, such large values of FF would not be experimentally feasible to realise with current nanofabrication techniques for such periods, which exhibit reduced repeatability of reactive ion etching for features $<50$ nm.

To realise a photonic Jackiw-Rebbi (JR) edge state, the band gap must be fully closed and re-opened to change the topological phase \cite{Lee2021}. Such \ch{WS2} gratings alone are therefore not suitable for this task, which lead to the design and fabrication of the inverted gratings as detailed in the main text.

\newpage
\section{Supplementary Note 3}
\label{sec:SI3}

\begin{figure*}[ht!]
\centering
  \includegraphics[scale=0.625]{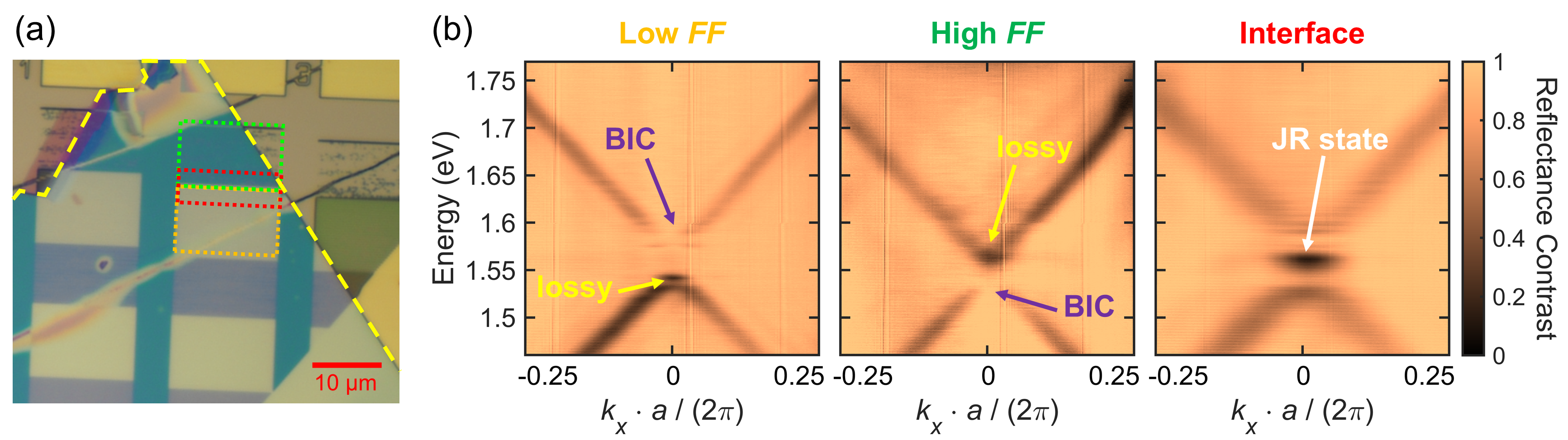}
\caption{\textbf{Experimental angle-resolved reflectance contrast measurements of an additional \ch{WS2} inverted double grating sample on gold. (a)} Optical microscope image of the structure with transferred top \ch{WS2} slab of thickness $t_\mathrm{slab}=44$ nm outlined by the yellow dashed line. Grating layer beneath is of thickness $t_\mathrm{gr}=52$ nm. Low FF grating denoted by the orange dotted box with period and filling factor $a_2=272$ nm, $F_2=0.84$ respectively. High FF grating denoted by the green dotted box with period and filling factor $a_1=356$ nm, $F_1=0.24$ respectively. Red dotted box corresponds to the grating interface region. \textbf{(b)} From left to right: angle-resolved reflectance contrast taken from the low FF grating, high FF grating, and interface region respectively. Clear band inversion is observed with an interface-localised JR state at the centre of the band gap.}
 \label{fig:SI3}
\end{figure*}

We present here another inverted grating sample, separate to those of the main text, to prove the repeatability of obtaining JR interface states via our transfer fabrication method. An optical microscope image of the sample is presented in Figure \ref{fig:SI3} (a), with the top \ch{WS2} slab outlined in yellow. The respective thicknesses of the grating and bulk layers are $t_\mathrm{gr}=52$ nm and $t_\mathrm{slab}=44$ nm. The low FF grating is outlined by the orange dotted box in Figure \ref{fig:SI3} (a), corresponding to the region of collected signal using the variable aperture in our Fourier setup (Figure 3 (d) of the main text). The same size collection region is used for the low and high FF gratings (green dotted box) to collect as much signal as possible. The gratings have period and filling factors $a_2=272$ nm, $F_2=0.84$ for the low FF grating, and $a_1=356$ nm, $F_1=0.24$ for the high FF grating respectively.

We subsequently measure the angle-resolved reflectance contrast for the three different regions as in the main text, and plot the results in Figure \ref{fig:SI3} (b). The left panel corresponds to signal from the low FF grating with a BIC on the upper energy branch and lossy mode on the lower branch. By tuning to high filling factor as in the central panel, there is clear inversion of the bands with the BIC on the lower energy branch. In this structure, the low and high FF grating band gaps are positioned at lower energies than the structure measured in Figure 4 of the main text, at around $1.55$ eV. By measuring at the interface region we observe a clear state in the centre of the band gap, which we attribute to a topologically-protected JR interface state. Here the variable aperture used was smaller than for the separate gratings (red dotted box in Figure \ref{fig:SI3} (a)) to focus on signal from the JR state at the interface. This real-space confinement results in a slight broadening of the modes in reciprocal space and energy owing to the Heisenberg uncertainty principle.

This additional structure shows the ability to precisely and repeatably fabricate inverted double grating structures from van der Waals materials to achieve JR interface states at different energies depending on the grating parameters and flake thicknesses chosen.

\newpage
\section{Supplementary Note 4}
\label{sec:SI4}

\begin{figure*}[ht!]
\centering
  \includegraphics[scale=1]{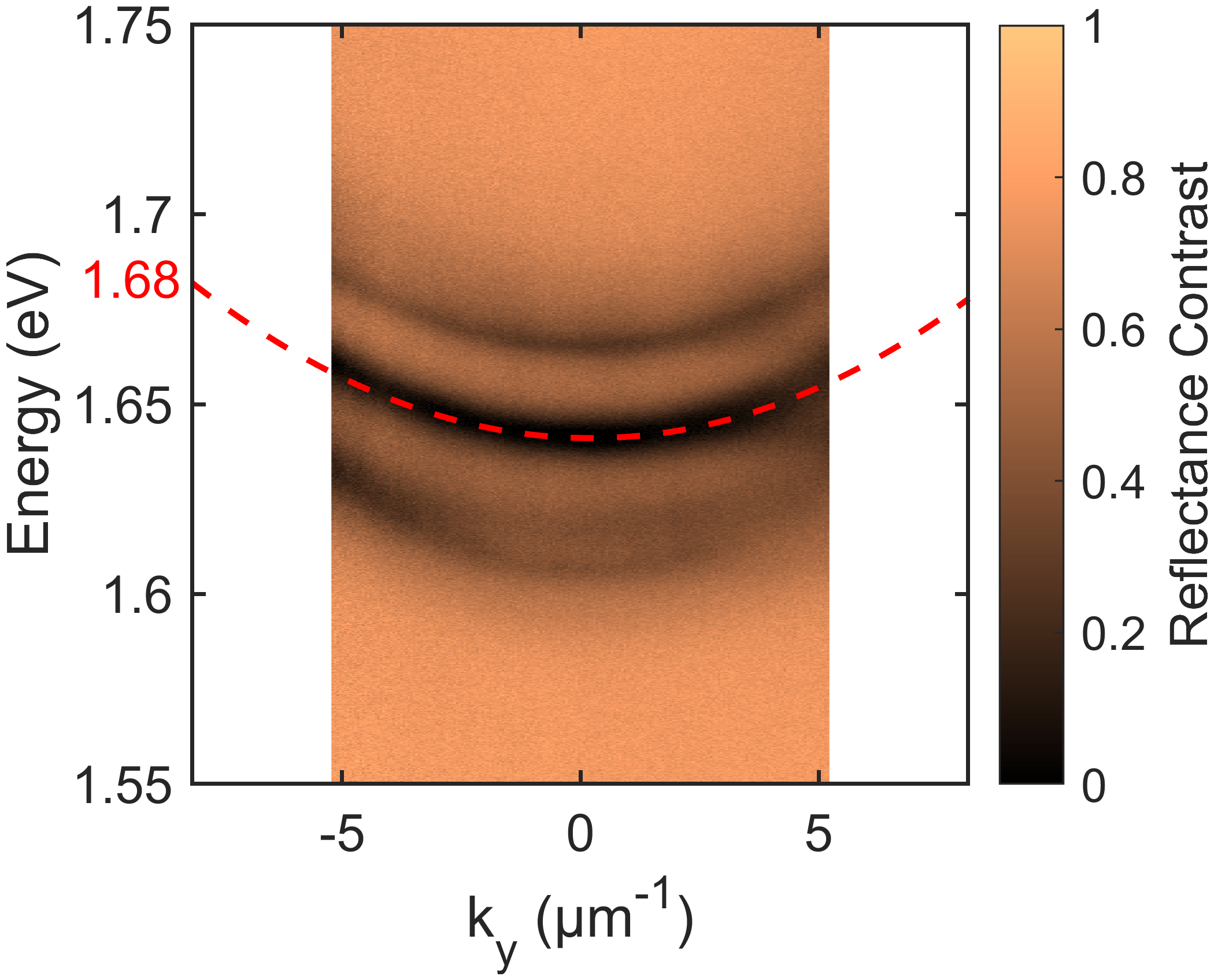}
\caption{\textbf{Extrapolated experimental angle-resolved reflectance contrast of interface region between \ch{WS2} double inverted gratings on gold along $k_y$ direction.} Upper and lower parabolic modes correspond to the grating lossy modes. Edge state dispersion fitted with a second order polynomial as shown by the red dashed line. Data extrapolated up to a wave vector corresponding to $60\degree$ incident angle.}
 \label{fig:SI4}
\end{figure*}

In order to accurately determine the correct excitation energy for the Jackiw-Rebbi edge state in \ch{WS2} double inverted gratings on gold with our scattering-type scanning near-field optical microscopy (s-SNOM) system, we extrapolated the measured dispersions from the middle panel of Figure 4 (b) of the main text. This data corresponds to angle-resolved reflectance contrast from the interface region between the high and low filling factor gratings, up to an angle of $\pm44\degree$ owing to our $0.7$ numerical aperture objective. For the s-SNOM measurements, the incident angle was fixed at $60\degree$ to the sample normal (i.e. tip axis), and so beyond our measured range with the Fourier setup. We therefore fitted the edge state dispersion to a second order polynomial as shown by the red dashed line in Figure \ref{fig:SI4}, and extrapolated up to higher wave vectors in $k_y$ (i.e. parallel to the grating axis) corresponding to an incident excitation angle of $60\degree$. The predicted JR state energy at this angle was $1.68$ eV, which agrees perfectly with our simulations from Figure 5 (b) of the main text. In addition, performing s-SNOM scans at this energy with light propagating along the grating axis showed field localisation at the grating interface, exactly where we expect the JR state to exist as in Figures 5 (d) and (e) of the main text.

\newpage
\section{Supplementary Note 5}
\label{sec:SI5}

\begin{figure*}[ht!]
\centering
  \includegraphics[scale=1.2]{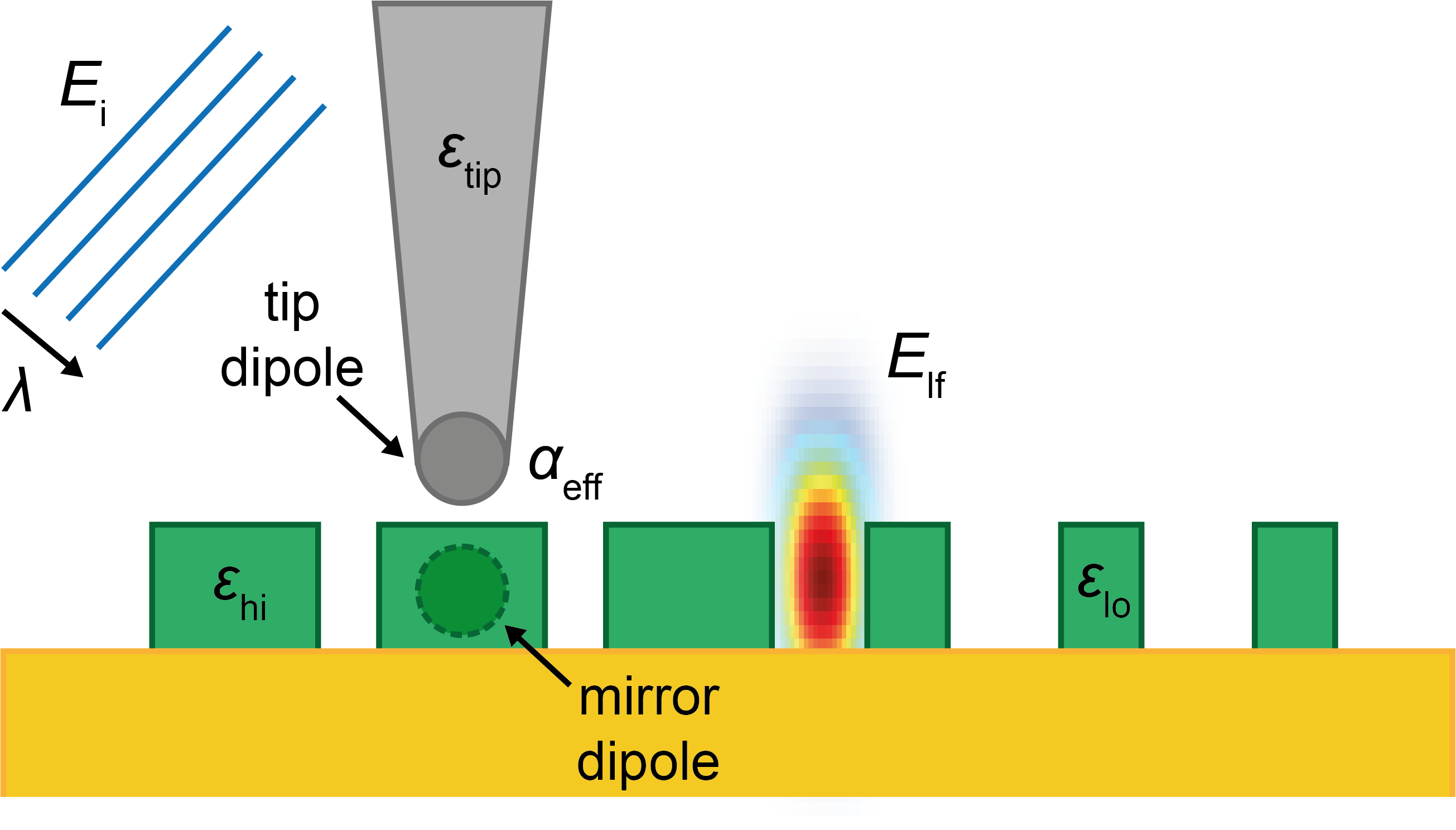}
\caption{\textbf{Schematic of s-SNOM probing regimes over different grating structures.} Tip dipole excited under illumination with incident field $E_\mathrm{i}$, leading to corresponding mirror dipole induced within the sample. Metallic tip has permittivity $\epsilon_\mathrm{tip}$, and the high and low filling factor gratings having effective permittivities $\epsilon_\mathrm{hi,lo}$ respectively. $\alpha_\mathrm{eff}$ represents the effective polarisability of the whole tip-sample interaction region. $E_\mathrm{lf}$ corresponds to the strength of local electric fields.}
 \label{fig:SI5}
\end{figure*}

Since s-SNOM works via measuring the scattering intensity from the tip-sample interaction region, the measured signal is dependent upon the refractive index of the material beneath the tip at any one time. The well-known point-dipole model \cite{keilmann2004near} considers a dipole excited at the apex of a metallic tip under incident illumination $E_\mathrm{i}$, as depicted schematically in Figure \ref{fig:SI5}. The tip dipole subsequently induces a mirror dipole within the sample beneath, with properties that depend on the sample permittivity. An overall effective polarisability of the tip-sample interaction region can thus be defined $\alpha_\mathrm{eff}$, which depends on both the permittivity of the tip and sample. The resulting scattering from this region can be expressed as $E_\mathrm{s}\propto\alpha_\mathrm{eff}E_\mathrm{i}$ \cite{keilmann2004near}, where $E_\mathrm{s},E_\mathrm{i}$ are the electric fields of the scattered signal and incident light respectively. As a result, the high and low filling factor gratings will have different baseline scattering intensities owing to their different effective permittivities, as illustrated in Figure \ref{fig:SI5}. Therefore, the total measured signal will be a combination of the intrinsic material properties of the sample, and the effect of the tip's interaction with locally confined electric fields $E_\mathrm{lf}$, e.g. from grating and topological resonances present. To minimise the material response, we normalise the scattered signal from each different grating separately, as done in Figures 5 (d) and (e) of the main text, enabling clearer comparison of the local electric fields intensities.

There are also other processes involved, such as scattered photons from the edge of the top \ch{WS_2} slab. This region is highlighted by the white dashed line in Figure 5 (d) of the main text, where we observe straight wavefronts propagating perpendicular to the diagonal edge of the slab. We attribute this to interference between the tip-sample-scattered photons, and photons scattered from the edge of the top slab directly to the detector without interacting with the tip. The result is a pattern of dark and bright fringes that decay in intensity as the distance from the edge increases. This effect does not interact with the JR edge state directly, and simply forms either constructive or destructive interference depending on the phase mismatch with the edge-scattered photons.

\newpage
\section{Supplementary Note 6}
\label{sec:SI6}

\begin{figure*}[ht!]
\centering
  \includegraphics[scale=0.54]{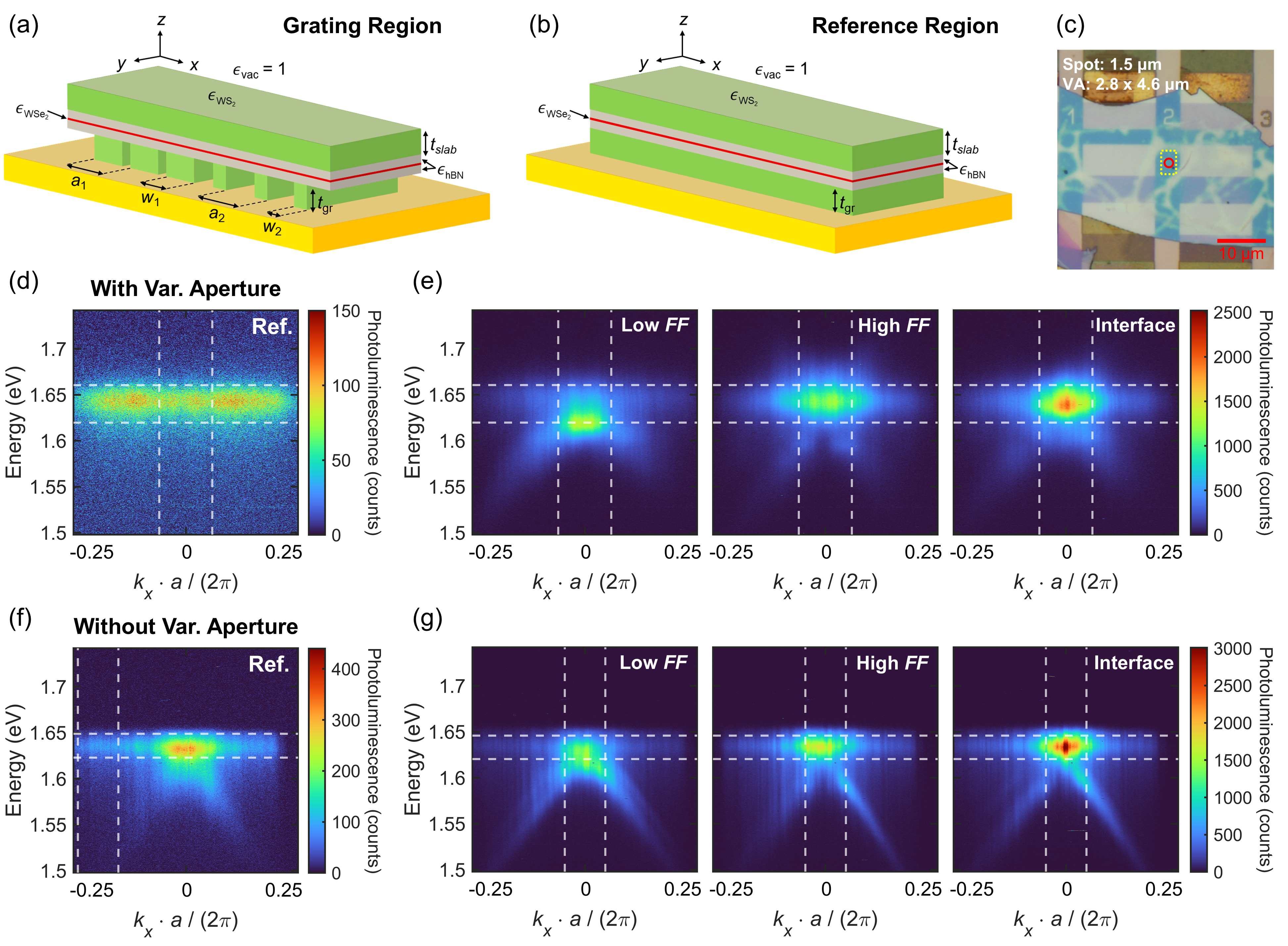}
\caption{\textbf{Angle-resolved photoluminescence measurements with and without variable aperture. (a)} Schematic of the grating heterostructure with patterned bottom \ch{WS2} creating a double grating interface. \textbf{(b)} Schematic illustrating the structure measured at the reference position consisting of an hBN-encapsulated monolayer \ch{WSe2} sandwiched between two unpatterned bulk \ch{WS2} slabs on a gold substrate. \textbf{(c)} Optical microscope image of grating heterostructure where the variable aperture and laser spot correspond to the yellow dotted box and red circle respectively. Relative sizes given at the top left in white. Laser spot centred over the reference region between gratings. \textbf{(d)} Angle-resolved photoluminescence (PL) taken from the reference region depicted in \textbf{(b)} with the variable aperture illustrated by the dotted yellow box in \textbf{(c)}. No dispersion is observed across $k_x$ with uniform PL signal around $1.64$ eV corresponding to the \ch{WSe2} excitonic emission. Dashed white lines highlight the integration region used to calculate PL directional enhancement from coupling to grating modes. \textbf{(e)} PL taken from the same positions as Figure 6 (d) of the main text with the variable aperture. Confinement in real space leads to the observed broadening of the peaks in reciprocal space and energy. \textbf{(f)} Angle-resolved PL taken from the reference region without the variable aperture. Signal is taken from the full imaged region in \textbf{(c)} leading to capturing emission from PL coupled to nearby gratings. Integration region is thus shifted left to avoid such interference. \textbf{(g)} PL taken from the same positions as Figure 6 (d) of the main text without the variable aperture. Larger collection region leads to narrower mode linewidths.}
 \label{fig:SI6}
\end{figure*}


Here we detail the method used to calculate the photoluminescence directional enhancement of monolayer \ch{WSe2} coupled to the JR interface state hosted within the grating heterostructure presented in Figure 6 of the main text. The basic principle is to first excite the structure with a $637$ nm laser with spot diameter $1.5$ \textmu m, and collect the resulting PL emission over a range of wave vectors $k_x$ using the Fourier spectroscopy setup (Figure 3 (d) of main text). A schematic of the grating heterostructure is shown in Figure \ref{fig:SI6} (a), where the excitation spot is moved perpendicular to the grooves (i.e. along $x$) with measurements taken at each point. We therefore collect PL coupled to both the low and high FF grating modes separately, and PL coupled to the JR state at the interface region depending on where the excitation/collection region is positioned. The PL at each $x$ position is then integrated over $k_x$ and energy, and divided by a constant reference measurement to give a directional enhancement factor of the PL. The reference region is depicted schematically in Figure \ref{fig:SI6} (b), and corresponds to a region with unpatterned bottom \ch{WS2} flake. We expect there to be no grating modes in this region and hence no coupling or modification of the PL intensity or dispersion. An optical microscope image of the measured grating heterostructure is shown in Figure \ref{fig:SI6} (c), where the red circle corresponds to the laser excitation spot positioned over the reference region between gratings. The dotted yellow box corresponds to the signal collection region taken using a rectangular variable aperture placed in the real-space plane as shown in Figure 3 (d) of the main text.

Use of the variable aperture ensured that signal only from the reference region was collected, and not leaked emission from the nearby gratings which were in close spatial proximity. The resulting angle-resolved reference PL is plotted in figure \ref{fig:SI6} (d), showing a uniform and flat dispersion at $\sim1.64$ eV as expected from monolayer \ch{WSe2}. Here there are no grating modes for the excitonic emission to couple to, and thus the PL is uniform in all directions. The three panels in Figure \ref{fig:SI6} (e) from left to right correspond to angle-resolved PL measurements taken via excitation and collection from the low FF grating, high FF grating, and interface region respectively. These positions are the same as for the spectra shown in Figure 6 (d) of the main text, however here we add the variable aperture in the collection path like with the reference measurement, ensuring signal only from each of the three regions was collected. We observe similar results to Figure 6 (d) of the main text, with coupling to the lossy mode of the low FF grating (left panel), a dip in PL intensity corresponding to the BIC of the high FF grating (centre panel), and enhanced directional emission via coupling to the JR state at the grating interface (right panel). Owing to the restricted collection region of $\sim2.8\times4.6$ \textmu m from the variable aperture, the resulting peaks in the PL spectra are noticeably broader in both wave vector and energy compared to without using the aperture. This is a result of the Heisenberg uncertainty relation between real space and momentum space. For Figure 6 (d) in the main text, we therefore opted to remove the variable aperture and collect from a much larger region (whole image area shown in Figure \ref{fig:SI6} (c)) to improve mode linewidths. We note that the excitation spot still remained small at $1.5$ \textmu m diameter to excite only local modes and not the whole sample.

The dashed white lines shown in all PL spectra of Figure \ref{fig:SI6} depict the integration regions used to calculate the PL directional enhancement factors. The aim was to quantify how well the JR interface state can enhance and direct coupled PL emission normal to the sample plane compared to PL from uncoupled monolayer. We thus chose to centre the integration region around the JR state as in the right panel of Figure \ref{fig:SI6} (e), with the size of the region corresponding to the respective linewidths in both $k_x$ and energy fitted via a Lorentzian curve. The PL directional enhancement was then simply the integrated PL over this region for each $x$ position measured over the gratings, divided by the same integrated region for the reference region PL.

The reference PL measurement taken without the variable aperture is plotted in Figure \ref{fig:SI6} (f), and was used for the PL directional enhancement factor calculations in Figure 6 (e) of the main text. Likely owing to light propagation through the crystal to nearby gratings, we capture some coupling of the \ch{WSe2} emission to the grating modes resulting in a weak parabolic mode dispersion. We therefore shift the integration region in $k_x$ to a position far from any dispersions, thus avoiding enhanced signal from grating mode coupling at the reference point. We note that this method is valid since the size of the integration region remains the same, and the true dispersion of the reference region was previously shown to be uniform with respect to $k_x$ in Figure \ref{fig:SI6} (d) using the variable aperture. Figure \ref{fig:SI6} (g) is the same as Figure 6 (d) of the main text but with the integration region highlighted for each panel for reference. There is clear enhancement of the directivity of monolayer \ch{WSe2} PL emission when coupled to the JR state at the grating heterostructure interface region, quantified as $22$ times stronger than PL from the same monolayer \ch{WSe2} with the same pump power ($216$ \textmu m) at the reference region, where the emission is uncoupled from any grating modes.

\newpage
\bibliography{SI}